\documentclass[a4paper,onecolumn,accepted=2023-01-20]{quantumarticle}
\pdfoutput=1
\usepackage[latin9]{inputenc}
\usepackage{epsfig}
\usepackage{verbatim}

\usepackage[normalem]{ulem}

\usepackage{ucs}
\usepackage{bbm}
\usepackage{orcidlink} 

\usepackage{hyperref}

\usepackage{amsmath}
\usepackage{amssymb}

\newcommand{\ket}[1]{|#1\rangle}

\newcommand{\be}{\begin{equation}}
\newcommand{\ee}{\end{equation}}
\newcommand{\eea}{\end{eqnarray}}
\newcommand{\bea}{\begin{eqnarray}}

\newcommand{\va}[1]{\ensuremath{(\Delta#1)^2}}
\newcommand{\vasq}[1]{\ensuremath{[\Delta#1]^2}}

\newcommand{\ex}[1]{\ensuremath{\langle{#1}\rangle}}
\newcommand{\exs}[1]{\ensuremath{\langle{#1}\rangle}}

\newcommand{\qed}{\ensuremath{\hfill \blacksquare}}

\newcommand{\kommentar}[1]{}

\newcommand{\forget}[1]{}
\newcommand{\tr}{\mbox{Tr}}

\newcommand{\EQ}[1]{Eq.~\eqref{#1}}
\newcommand{\EQS}[1]{Eqs.~\eqref{#1}}
\newcommand{\EQL}[1]{Equation~\eqref{#1}}
\newcommand{\SEC}[1]{Sec.~\ref{#1}}
\newcommand{\FIG}[1]{Fig.~\ref{#1}}
\newcommand{\TABLE}[1]{Table~\ref{#1}}

\newcommand{\REF}[1]{Ref.~\cite{#1}}
\newcommand{\REFS}[1]{Refs.~\cite{#1}}

\newcommand{\APP}[1]{Appendix~\ref{#1}}

\newcommand{\SUPERA}{^{a}}
\newcommand{\SUPERB}{^{b}}
\newcommand{\SUPERN}{^{s}}
\newcommand{\JTILDE}{\mathcal J}
\newcommand{\JLSUPERN}{j_l^{(n)}}
\newcommand{\WELL}{s}

\begin{document}

\title{Number-phase uncertainty relations and bipartite entanglement detection in spin ensembles}

\author{Giuseppe Vitagliano\,\orcidlink{0000-0002-5563-3222}}
\affiliation{Institute for Quantum Optics and Quantum Information (IQOQI), \\Austrian Academy of Sciences, AT-1090 Vienna, Austria}
\affiliation{Theoretical Physics, University of the Basque Country UPV/EHU, ES-48080 Bilbao, Spain}
\author{Matteo Fadel\,\orcidlink{0000-0003-3653-0030}}
\affiliation{Department of Physics, ETH Z\"urich, CH-8093 Z\"urich, Switzerland}
\affiliation{Department of Physics, University of Basel, CH-4056 Basel, Switzerland}
\author{Iagoba Apellaniz\,\orcidlink{0000-0002-3688-3242}}
\affiliation{Theoretical Physics, University of the Basque Country UPV/EHU, ES-48080 Bilbao, Spain}
\affiliation{EHU Quantum Center, University of the Basque Country UPV/EHU, \\Barrio Sarriena s/n, ES-48940 Leioa, Biscay, Spain}
\affiliation{Mechanical and Industrial Manufacturing Department, Mondragon Unibertsitatea, ES-20500 Mondrag\'on, Spain}
\author{Matthias~Kleinmann\,\orcidlink{0000-0002-5782-804X}}
\affiliation{Naturwissenschaftlich-Technische Fakult\"at, Universit\"at Siegen, DE-57068 Siegen, Germany}
\affiliation{Theoretical Physics, University of the Basque Country UPV/EHU, ES-48080 Bilbao, Spain}
\author{Bernd L\"ucke\,\orcidlink{0000-0002-4883-8756}}
\affiliation{Institut f\"ur Quantenoptik, Leibniz Universit\"at Hannover, DE-30167 Hannover, Germany}
\author{Carsten Klempt\,\orcidlink{0000-0003-2402-3162}}
\affiliation{Institut f\"ur Quantenoptik, Leibniz Universit\"at Hannover, DE-30167 Hannover, Germany}
\affiliation{Deutsches Zentrum f\"ur Luft- und Raumfahrt e.V. (DLR), Institut f\"ur Satellitengeod\"asie und Inertialsensorik,\\DLR-SI, Callinstra\ss e 36, DE-30167 Hannover, Germany}
\author{G\'eza T\'oth\,\orcidlink{0000-0002-9602-751X}}
\affiliation{Theoretical Physics, University of the Basque Country UPV/EHU, ES-48080 Bilbao, Spain}
\affiliation{EHU Quantum Center, University of the Basque Country UPV/EHU, \\Barrio Sarriena s/n, ES-48940 Leioa, Biscay, Spain}
\affiliation{Donostia International Physics Center (DIPC),  ES-20080 San Sebasti\'an, Spain}
\affiliation{IKERBASQUE, Basque Foundation for Science, ES-48013 Bilbao, Spain}
\affiliation{Institute for Solid State Physics and Optics, Wigner Research Centre for Physics, HU-1525 Budapest, Hungary}
\email{toth@alumni.nd.edu}
\homepage{http://www.gtoth.eu}

\begin{abstract}
We present a method to detect bipartite entanglement based on number-phase-like uncertainty relations in split spin ensembles. First, we derive an uncertainty relation that plays the role of a number-phase uncertainty for spin systems. It is important that the relation is given with well-defined and easily measurable quantities, and that it does not need assuming infinite dimensional systems. Based on this uncertainty relation, we show how to detect bipartite entanglement in an unpolarized Dicke state of many spin-$1/2$ particles. The particles are split into two subensembles, then collective angular momentum measurements are carried out locally on the two parts. First, we present a bipartite Einstein-Podolsky-Rosen (EPR)  steering criterion. Then, we present an entanglement condition that can detect bipartite entanglement in such systems. We demonstrate the utility of the criteria by applying them to a recent experiment given in K. Lange {\it et al.} [\href{http://doi.org/10.1126/science.aao2035}{Science {\bf 360}, 416 (2018)}]  realizing a Dicke state in a Bose-Einstein condensate of cold atoms, in which the two subensembles were spatially separated from each other.  Our methods also work well if split spin-squeezed states are considered. We show in a comprehensive way how to handle experimental imperfections, such as the nonzero particle number variance including the partition noise, and the fact that, while ideally BECs occupy a single spatial mode, in practice the population of other spatial modes cannot be fully suppressed.
\end{abstract}

\maketitle 

\section{Introduction}

Entanglement lies at the heart of many problems in quantum mechanics and has attracted an increasing attention in recent years \cite{Guhne2009Entanglement,Horodecki2009Quantum,Friis2019,Frerot2023Probing}. Entangled states can be used for metrology in order to obtain a sensitivity higher than the shot-noise limit \cite{Pezze2009Entanglement,Hyllus2012Fisher,Toth2012Multipartite} and can also be used as a resource for certain quantum information processing tasks \cite{Raussendorf2001A,Raussendorf2003Measurement-based,Gottesman1995Class,Cleve1999How,Curty2004Entanglement}. Entanglement also plays an important role in quantum computing making it possible that quantum computers outperform their classical counterparts for several problems such as prime factoring or searching \cite{Shor1999Polynomial-Time,Grover1996A}. Moreover, experiments realizing macroscopic quantum entanglement might give answers to fundamental questions in quantum physics \cite{Diosi1989Models,Frowis2012Measures}.

When in an experiment entanglement is created, it is important to detect it. Thus, in many quantum physics experiments the creation of an entangled state is followed by measurements. Based on the results of these measurements, the experimenters conclude that the produced state was entangled. However, in many-particle experiments the possibilities for quantum control are very limited. In particular, the particles cannot be individually addressed. In such systems, the entanglement can be created and detected with collective operations. The first entanglement criterion based on such collective quantities was the spin-squeezing criterion that detects entanglement in an ensemble of spin-$1/2$ particles, and it detects entangled states that are close to be fully polarized \cite{Sorensen2001Entanglement}, even multipartite entanglement can be detected in such states \cite{Sorensen2001Entanglement}. Later, a full set of generalized spin-squeezing criteria have also been defined \cite{Toth2007Optimal,Toth2009Spin}. Such criteria can detect states very different from usual spin-squeezed states. For instance, unlike spin-squeezed states, unpolarized symmetric Dicke states have a zero expectation value for all spin components and the precision in parameter estimation in linear interferometers using such states can reach, in principle, the maximal Heisenberg-scaling with the number of particles \cite{Lucke2011Twin,Toth2014Quantum,Pezze2018Quantum}. Even criteria for multiparticle entanglement have been developed for Dicke states \cite{Duan2011Entanglement,Lucke2014Detecting,Vitagliano2017Entanglement} and there are criteria that detect metrologically useful entanglement \cite{Zhang2014Quantum,Apellaniz2015Detecting,Apellaniz2017Optimal}. Moreover, criteria for an ensemble of particles with a spin larger than $1/2$ have also been found \cite{Duan2002Quantum,Vitagliano2011Spin,Vitagliano2014Spin,Toth2010Generation}.

The entanglement criteria mentioned above have been tested in a variety of physical systems \cite{Ma2011Quantum,Toth2014Quantum,Pezze2018Quantum}.  In photonic systems, symmetric Dicke states have been created \cite{Kiesel2007Experimental,Wieczorek2009Experimental,Prevedel2009Experimental,Krischek2011Useful,Chiuri2012Experimental}. In cold trapped ions, spin-squeezed states \cite{Bohnet2016Quantum}, W-states, i.e., Dicke states with a single excitation \cite{Haffner2005Scalable} have been realized and there is a general scheme for obtaining various kind of Dicke states \cite{Hume2009Preparation}. In atomic ensembles, spin-squeezed states \cite{Gross2012Spin,Wasilewski2010Quantum,Fernholz2008Spin,Hald1999Spin,Julsgaard2001Experimental,Hammerer2010Quantum} and many-body singlet states have been realized \cite{Behbood2014Generation,Kong2020Measurement-induced}. Large scale quantum entanglement has also been created in Bose-Einstein Condensates (BEC). Spin-squeezed states  of two-state atoms have been realized in several experiments \cite{Gross2012Spin,Esteve2008Squeezing,Gross2010Nonlinear,Riedel2010Atom-chip-based,Ockeloen2013Quantum,Muessel2014Scalable,Pezze2018Quantum}.  Multipartite entanglement has been detected in these states, and it has also been demonstrated that such states are useful for quantum metrology.

From the point of view of our paper it is very relevant that, recently, symmetric Dicke states have also been created in BECs \cite{Lucke2011Twin,Hamley2012Spin-nematic}. Apart from entanglement, even metrological usefulness of Dicke states has been verified \cite{Lucke2011Twin,Apellaniz2015Detecting}. A Dicke state of around 8000 particles has been realized and multipartite entanglement up to  $28$ particles has been detected  within an error of two standard deviations \cite{Lucke2014Detecting,Vitagliano2017Entanglement}. Moreover, Dicke states of more than 10000 spin-1 atoms and 630-particle entanglement has been detected within an error of a single standard deviation \cite{Zou2018Beating}.  

As a natural next step, it is important to detect entanglement between two parts of such collective spin states, since bipartite entanglement between spatially separated subsystems is the type of entanglement most useful for quantum information processing applications. It has been shown that multipartite entanglement of bosonic two-state atoms can be converted into bipartite entanglement by splitting the ensemble into two \cite{Killoran2014Extracting}.  However, can this powerful theoretical result be verified in an experiment, where noise and other imperfections are present? The question is even more timely since the detection of bipartite entanglement of high-dimensional systems raised a lot of attention \cite{Krenn6243Generation,Erker2017Quantifying}. In a recent experiment, bipartite entanglement was successfully detected in a Dicke state created with thousands of atoms and distributed into two spatially separated regions \cite{Lange2018Entanglement}.

Another form of quantum correlations, stronger than entanglement, termed Einstein-Podolsky-Rosen (EPR) steering \cite{Einstein1935Can,Reid1989Demonstration,Reid2001The,Reid2009Colloquium,Cavalcanti2009Experimental}, has also been recently detected in photonic systems \cite{Ou1992Realization}, as well as in BECs \cite{Peise2015Satisfying}. Methods to detect EPR correlations in spatially separated parties, i.e., in a two-well BECs  have also been studied theoretically \cite{He2011Einstein,Jing2019Split,Guo2021Detecting}.  Experimentally, the observation of EPR correlations between spatially separated modes was achieved in split-squeezed atomic ensembles \cite{Fadel2018Spatial}, while entanglement has also been detected using a criterion developed by Giovannetti {\it et al.} in \REF{Giovanetti2003Characterizing}. EPR correlations have also been detected with spin-nematic squeezing \cite{Kunkel2018Spatially}.

All entanglement conditions employed in the recent experiments mentioned above are based on variances of collective spin components, and it was recently observed how such criteria can be converted into lower bounds to entanglement monotones, thus allowing for the quantification of entanglement with the same information~\cite{FadelVitagliano_2021}. Thus, finding efficient entanglement witnesses is very relevant also for the quantification of entanglement.

In this paper, we present entanglement criteria that verify the presence of bipartite entanglement between spatially separated parts of a condensate. We also introduce an EPR steering criterion.  Our criteria are particularly well suited for symmetric unpolarized Dicke states given as
\be
\vert {\rm D}_N \rangle=\binom{N}{N/2}^{-1/2} \sum_k \mathcal{P}_k (\vert 1\rangle^{\otimes N/2}\vert 0\rangle^{\otimes N/2}),\label{eq:Dickestate}
\ee
where $\ket0$ and $\ket1$ are the basis states of a two-level atom, and $N$ is even and the summation is over all distinct permutations of the particles. We will call $\vert {\rm D}_N \rangle$ simply Dicke state in the following. For an equal partitioning of the $N$ particles, the entanglement of formation of $\vert {\rm D}_N \rangle$ is large, which will be discussed in \APP{sec:DickeState}. This gives us a further strong motivation to try to detect bipartite entanglement in these states \cite{Stockton2003Characterizing}. Our methods also work for other quantum states, such as spin-squeezed states.

\begin{figure}[t]
\begin{center}
\epsfxsize7.1cm \epsffile{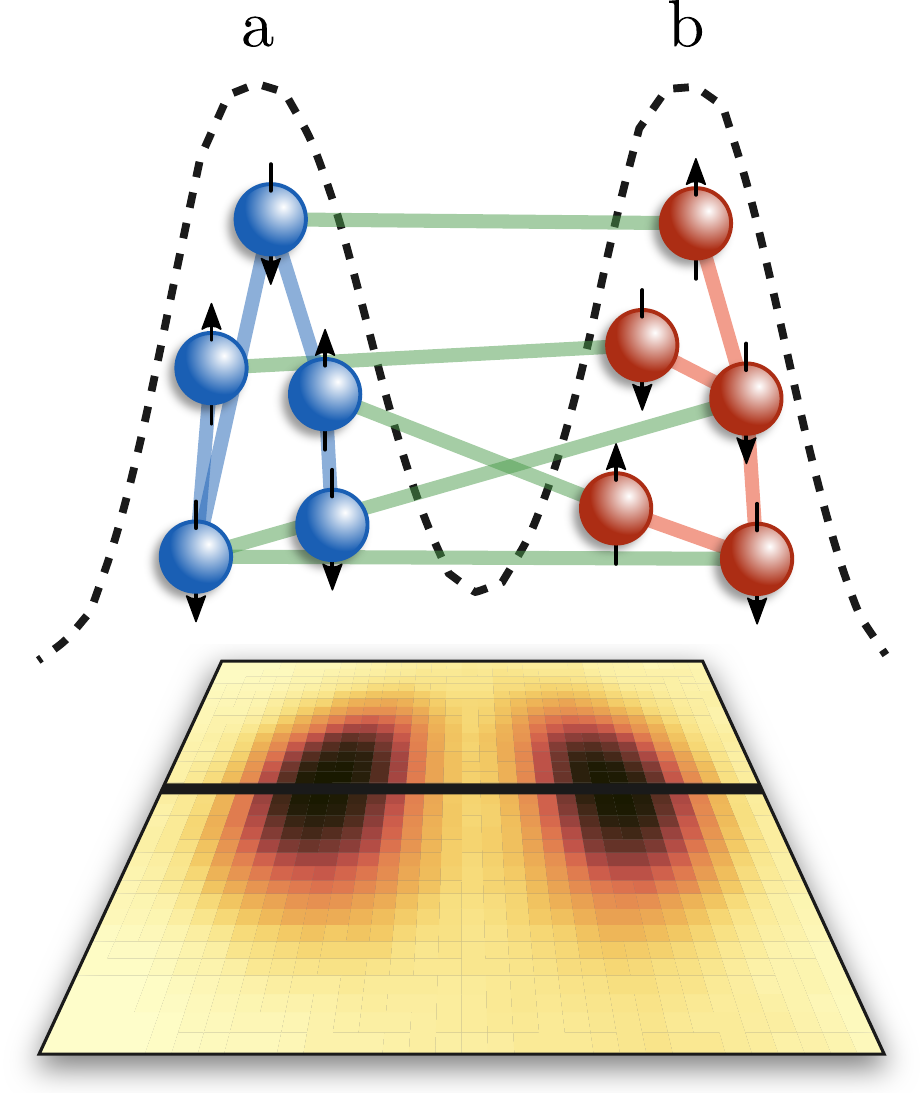}\vskip0.0cm
\caption{Entanglement detection between two spatially separated ensembles of spin-$1/2$ particles. The atomic density profile obtained by averaging 3329 measurement results in the experiment presented in \REF{Lange2018Entanglement} is shown.}
\label{fig:doube_well}
\end{center}
\end{figure}

We obtain criteria that detect entanglement and EPR steering between two parts of the Dicke state in Eq.~\eqref{eq:Dickestate}. For that, we divide the system into two subsystems, which we denote by ``$a$'' and ``$b.$'' The particle number operators in the two subsystems fulfill
\be
N_a+N_b=N.\label{eq:NabN}
\ee
In practice, this means that an atomic cloud is spatially separated into two regions as in \FIG{fig:doube_well}.  At this point we have to stress an important property of real experiments. The total particle number fluctuates from experiment to experiment. Moreover, even if the total particle number remained fixed,  the particle numbers $N_a$ and $N_b$ will change from experiment to experiment due to fundamental quantum effects that we will discuss later. Thus, the quantum states do not live in a single Hilbert space with a fixed $N_a$ and $N_b,$ but rather they live in several such Hilbert spaces. A general state of our system can be written as \cite{Hyllus2010Entanglement,Hyllus2012Entanglement}
\be
\varrho=\sum_{j_a,j_b} p_{j_a,j_b} \varrho_{j_a,j_b},\label{eq:dmat_varNnonzero}
\ee
where $\varrho_{j_a,j_b}$ have definite particle numbers in the two subsystems, and $p_{j_a,j_b}$ are probabilities. Here, the local total angular momentum depends on the local particle number via the relation $j_s=N_s/2$ for $s=a,b.$  

Note that here we are discarding superpositions between states with different number of particles within subsystems $a$ and $b.$  The operators appearing in our entanglement criteria conserve the particle number within the subsystems, thus they cannot distinguish superpositions of states with different particle numbers from a mixture of such states. In any case, even if the state of the system had a superposition of states with different subsystem  particle numbers, we can remove them with the projection
\be
\varrho_{j_a,j_b}=P_{j_a,j_b} \sigma P_{j_a,j_b}, \label{eq:dmat_varNnonzero2}
\ee
where $\sigma$ is the state of the system and $P_{j_a,j_b}$ projects to a subspace with fixed particle numbers \cite{Hyllus2012Entanglement,Toth2004Entanglement}. Since our criteria cannot distinguish $\sigma$ from $\varrho$ in \EQ{eq:dmat_varNnonzero}, we will consider states of the form given in \EQ{eq:dmat_varNnonzero}. This is also the relevant case in experiments, where the measurement will always project onto a state with a definite particle number.

Our criteria are formulated with the moments of the collective angular momentum coordinates. For a state with a given $j_a$ and $j_b,$ they are defined as
\be
J_l=\sum_{n=1}^N \JLSUPERN,\label{eq:coll}
\ee
where $l=x,y,z,$ and $\JLSUPERN$ denotes a component of the $n^{\rm th}$ spin. Analogously, we can define $J_l^a$ and $J_l^b$ for the two subsystems. Naturally, the angular momentum components fulfill
\be
-j_s\openone\le J_l^s\le j_s\openone.
\ee.

For states given in \EQ{eq:dmat_varNnonzero}, we compute expectation values of observables as follows. Let us consider an operator 
\be
O=A f(\hat{j}_a,\hat{j}_b),\label{eq:O}
\ee
where we used the notation ``$\;\hat{}\;$'' stressing that $\hat{j}_a,\hat{j}_b$ are the total angular momentum operators, and are defined as $\hat{j}_a=\hat{N}_a/2$  and $\hat{j}_b=\hat{N}_b/2,$ where $\hat{N}_a$ and $\hat{N}_b$ are particle number operators.  For simplicity, we will omit ``$\;\hat{}\;$'' in the future. It will be clear from the context, when we talk about operators. Moreover, $f(x,y)$ denotes a two-variable function and $A$ is an operator constructed as a function of angular momentum components $J_l, J_l^a$ and $J_l^b.$ Then, the expectation value of $O$ can be obtained as
\be
\exs{ O}_{\varrho}=\sum_{j_a,j_b} p_{j_a,j_b} \ex{A}_{\varrho_{j_a,j_b}} f(j_a,j_b).\label{eq:exvarn}
\ee

For subsystem $a$, a formalism similar to that of \EQ{eq:dmat_varNnonzero} can be used
\be
\varrho_a=\sum_{j_a} p_{j_a} \varrho_{j_a}^a,\label{eq:dmat_varNnonzero_a}
\ee
where $\varrho_{j_a}$ have definite particle numbers. The expectation value of an operator $O^a,$ $\ex{O^a}_{\varrho_a}$ can be computed analogously to \EQ{eq:exvarn}. Analogous statements hold for subsystem $b.$

In this paper, we will present criteria that are satisfied by a {\it local hidden state model}. A local hidden state model means the following. There is a hidden variable $\lambda,$ with a probability distribution given by $p_{\lambda}.$ For each value of the hidden variable $\lambda,$ there is a (hidden) state of subsystem $a$ described by the normalized density matrix $\varrho^a_\lambda.$ Any expectation value of an operator $O^a$ acting on $a$ can be obtained as a weighted average of expectation values for a given $\lambda$ 
\be
\ex{O^a}=\sum_{\lambda} p_{\lambda} {\rm Tr}(\varrho^a_\lambda O^a).\label{eq:Qa}
\ee
Subsystem $b,$ on the other hand, is described differently.  The local hidden state model provides for each $\lambda$ and operator $O^b$ acting on subsystem $b$ a distribution $p(x\vert O^b, \lambda),$ where $x$ are the outcomes of measuring $O^b.$ If the correlations cannot be described by such a  local hidden state model, then $\varrho$ is called {\it steerable}. We will give a more complete description in the main text.

We will also present criteria to detect entanglement in the system. The state $\varrho$ given in \EQ{eq:dmat_varNnonzero} is separable if and only if all $\varrho_{j_a,j_b}$ are separable, i.e., they are mixtures of product states~\cite{Hyllus2012Entanglement} and can be written as 
\be
\varrho_{j_a,j_b}=\sum_k p_k \varrho_{j_a,j_b,k}^{a} \otimes \varrho_{j_a,j_b,k}^{b}, \label{eq:sep}
\ee
where $\varrho_{j_a,j_b,k}^{a}$ and $\varrho_{j_a,j_b,k}^{a}$ are states of the subsystems $a$ and $b$, respectively. 
If a quantum state $\varrho_{j_a,j_b}$ cannot be decomposed as in \EQ{eq:sep}, then it is entangled and the mixture of such states as in \EQ{eq:dmat_varNnonzero} is also entangled.

We will now present the three inequalities that will be proven in \SEC{sec:ourcriteria}. First, we present the basis of our work, an uncertainty relation that can be considered a number-phase uncertainty for atomic ensembles. Unlike in the case of bosonic modes, now we do not assume infinite systems. It is also important that our relation is given in terms of well defined and easily accessible quantities.

{\bf Observation 1.} For any quantum state,  the following uncertainty relation
\be
	\left( (\Delta J_z^s)^2 + \frac 1 4 \right)
		\frac {(\Delta J_x^s)^2 + (\Delta J_y^s)^2}
			{\langle (J_x^s)^2 \rangle + \langle (J_y^s)^2 \rangle}
	\ge \frac 1 4 \label{eq:unc}
\ee
holds for $s=a,b.$ Here, we formulated the uncertainty relation for one of the two subsytems.

Based on this, we find the following EPR steering condition. Essentially, a local hidden state model means the following. Subsystem $a$ is described by a density matrix conditioned on measurements on subsystem $b.$ Thus, criteria satisfied by states with a local hidden state model are uncertainty relations for subsystem $a$ conditioned on measurements on subsystem $b.$

{\bf Observation 2.} All quantum states that admit a local hidden state model must satisfy
\bea
&& \left[(\Delta_{\rm inf} J_z\SUPERA)^2+\frac{1}{4}\right] 
\left[(\Delta_{\rm inf} J_x\SUPERA)^2 + (\Delta_{\rm inf} J_y\SUPERA)^2\right] \ge \frac1 4
\left\langle \frac{(J_x\SUPERA)^2 + (J_y\SUPERA)^2 }{\sqrt{j_{a}(j_{a}+1)}} \right\rangle^2.
\label{eq:stcond}
\eea
Here, the inference variance $(\Delta_{\rm inf} J_l\SUPERA)^2$ is the variance of estimating $J_l\SUPERA$ based on measurement results on subsystem $b$ for quantum states with a local hidden state model. Note that it can be smaller than $(\Delta J_l\SUPERA)^2.$ When the criterion is applied to an experiment, one considers the substitution
\be
(\Delta_{\rm inf} J_l\SUPERA)^2 \rightarrow [\Delta( J_l\SUPERA-J_{l,{\textrm{est}}}\SUPERA)] ^2,\label{eq:INF}
\ee
where the estimates $J_{l,{\rm est}}\SUPERA$ are given as 
\be
J_{l,{\rm est}}\SUPERA= - g_lJ_l\SUPERB , 
\label{eq:est}
\ee
for $l=x,y,z,$ with $g_l$ constants\footnote{{Instead of \EQS{eq:INF} and \eqref{eq:est}, typically they write \cite{Reid1989Demonstration,Reid2001The,Reid2009Colloquium,Cavalcanti2009Experimental}
\be
(\Delta_{\rm inf} J_l\SUPERA)^2=\langle( J_l\SUPERA+g_l J_l\SUPERB+d_l) ^2\rangle,
\ee
where $d_l$ is a real number.  We have to choose $d_l$ such that the variance is minimal. This is the case if $d_l=-\ex{J_l\SUPERA+g_l J_l\SUPERB}.$ With that choice,  $\langle( J_l\SUPERA+g_l J_l\SUPERB+d_l) ^2\rangle=[\Delta( J_l\SUPERA+g_l J_l\SUPERB)] ^2$ holds, and we arrive at \EQ{eq:INF}.}}. Any quantum state that violates Eq.~(\ref{eq:stcond}) is steerable. The inequality in \EQ{eq:stcond} holds for all non-steerable states, and for any choice of the real parameters $g_l.$ Therefore, the latter can be optimized in such a way that the violation is maximized, i.e., the left-hand side of  \EQ{eq:stcond} is minimized. Later, we will show that for an ideal Dicke state of many particles, the optimal choice for $g_l$ is 
\be
g_x=g_y=-1,\quad g_z=1. \label{eq:glopt}
\ee

Then, we present a criterion detecting entanglement.

{\bf Observation 3.} For separable states in bipartite systems 
\be   
\left[\va{J_z}+\frac 1 4 \right] \left[ \va{J_x^-}+\va{J_y^-} \right] 
\ge 
\left\langle \frac{J_x^2+J_y^2}{\sqrt{N(N+2)}} \right\rangle^2\label{eq:rightJx2Jy2B_withoutsqrtB}
\ee
holds, where we define the difference between the angular momentum components in the two parts as
\be
J_l^{-}=J_l\SUPERA-J_l\SUPERB \label{eq:Jlpm}
\ee
for $l=x,y.$ 

If we set the $g_l$ according to \EQ{eq:glopt} then the left-hand side of \EQ{eq:stcond} is identical to the left-hand side of \EQ{eq:rightJx2Jy2B_withoutsqrtB}. Let us compare the  right-hand sides. For states living in the symmetric (i.e., bosonic) subspace such that $J_z\SUPERA, J_z\SUPERB \approx 0,$ as we will see later, the right-hand side of \EQ{eq:stcond} is around 4 times smaller than that of \EQ{eq:rightJx2Jy2B_withoutsqrtB}. Thus, for such states, it is more difficult to violate the EPR steering criterion \eqref{eq:stcond} than the entanglement criterion \eqref{eq:rightJx2Jy2B_withoutsqrtB}, which is consistent with the fact that steering is typically harder to get than entanglement.

In this paper, we will prove the relations mentioned above and show that these relations detect entanglement in a condensate split into two parts. Our method will handle the problem of fluctuating particle numbers, which has two different manifestations. On the one hand, the total particle number varies from experiment to experiment. On the other hand, when the ensemble is split, the particle number in the two subensembles will not be exactly half of the total particle number. Both of these effects make the detection of entanglement more difficult, since they decrease the violation of the entanglement criteria. An appropriate normalization of the measured quantities  can reduce the problem mentioned above. We show this in detail for the case of the splitting noise, which we call partition noise.

Our method can also tolerate another imperfection appearing in experiments. While ideally BECs occupy a single spatial mode, in practice the population of other spatial modes cannot be fully suppressed. This is partly due to the fact that BECs are prepared experimentally at some nonzero temperature. Hence, a quantum state realized with a BEC of two-level atoms will never be in a perfectly symmetric state \cite{Hyllus2012Entanglement}. Due to the complexity of the appropriate modelling, such effects have not been considered in many cases when the state of the ensemble was obtained via tomography, and in some cases they were taken into account through ad-hoc methods. We work out criteria that are valid also for states that are not perfectly symmetric. 

Our paper is organized as follows. In \SEC{sec:ourcriteria}, we prove the uncertainty relation (\ref{eq:unc}). We also prove the EPR steering criterion (\ref{eq:stcond}) and the entanglement criterion (\ref{eq:rightJx2Jy2B_withoutsqrtB}). In  \SEC{sec:relevant quantites}, we calculate relevant quantities for the split Dicke state and present other versions of our criteria with a normalization that handles the partition noise in Dicke states, i.e., the fluctuation of the particle numbers in the two halves of the system due to splitting. In \SEC{sec:appl}, we apply our criteria to various ideal quantum states as well as to an experiment in which a Dicke state has been prepared. Furthermore, we also test our criteria numerically on split spin-squeezed states.

\section{Derivation of the inequalities}

\label{sec:ourcriteria}

In this section, we prove the  number-phase-like uncertainty relation given in Observation~1. Then, we prove the EPR relation given in Observation~2 and the entanglement criterion in  Observation~3.

\subsection{Number-phase-like uncertainty}

Here, we derive an uncertainty relation that resembles a number-phase complementarity relation.

{\it Proof of Observation 1.} Let us first prove that the usual Heisenberg uncertainty relation for some $A$ and $B$ operators
\bea
(\Delta A^s)^2 (\Delta B^s)^2&\ge& \frac1 4 \ex{C^s}^2 \label{eq:Heisenberg}
\eea
with $C^s=i[A^s,B^s]$ is valid for a fluctuating number of particles and we formulated the uncertainty for one of the subsystems $s=a,b.$ Let us consider the following series of inequalities
\bea
        (\Delta A^s)^2 (\Delta B^s)^2 &\ge& \left(\sum_{j_s} p_{j_s} (\Delta A^s)^2_{\varrho_{j_s}}\right) \left( 
		\sum_{j_s} p_{j_s} (\Delta B^s)^2_{\varrho_{j_s}}\right) \nonumber \\
	&\ge&\frac1 4 \left(\sum_{j_s} p_{j_s} \vert \ex{C^s}_{\varrho_{j_s}}\vert \right)^2 \ge\frac1 4 \ex{C^s}^2,
	\label{eq:uncA}
\eea
where again $s=a,b.$ The first inequality in \EQ{eq:uncA} is due to the concavity of the variance. The second inequality holds because \EQ{eq:Heisenberg} is valid for a state with fixed $j_s.$ It is also due to the relation (see, e.g., \REF{Toth2022Uncertainty}) 
\begin{equation} \label{eq:ineq}
\left(\sum_k p_k a_k \right)\left(\sum_k p_k b_k \right)\ge\left(\sum_k p_k \sqrt{a_k b_k}\right)^2,
\end{equation}
where $a_k,b_k\ge0,$ which can be obtained straightforwardly from the Cauchy-Schwarz inequality\footnote{The Cauchy-Schwarz inequality is $\left(\sum_{k=1}^n u_k v_k\right)^2 \leq \left(\sum_{k=1}^n u_k^2\right) \left(\sum_{k=1}^n v_k^2\right).$ \EQL{eq:ineq} can be proved taking $u_k=\sqrt{p_k a_k}$ and $v_k=\sqrt{p_k b_k}.$}. The last inequality in \EQ{eq:uncA} is due to the basic properties of the absolute value. With that we proved that the Heisenberg uncertainty relation is true for states with a varying particle number. Note that our proof is valid even if $A^s$ and $B^s$ depend on $j_s$.

Now we consider the Heisenberg uncertainty relations 
\bea
(\Delta J_z^s)^2 (\Delta J_x^s)^2&\ge& \frac 1 4 \langle J_y^s \rangle ^2,\nonumber\\
(\Delta J_z^s)^2 (\Delta J_y^s)^2&\ge& \frac 1 4 \langle J_x^s \rangle ^2.
\eea
By summing the two inequalities we obtain 
\bea
(\Delta J_z^a)^2 [(\Delta J_x^s)^2+(\Delta J_y^s)^2]&\ge& \frac 1 4 (\langle J_x^s \rangle ^2+\langle J_y^s \rangle ^2).
\eea
Then, by adding $[(\Delta J_x^s)^2+(\Delta J_y^s)^2]/4$ to both sides, and dividing by ${\langle (J_x^s)^2 \rangle + \langle (J_y^s)^2 \rangle},$ we arrive at the uncertainty relation \EQ{eq:unc}. $\qed$

\EQL{eq:unc} is similar to a number-phase uncertainty relation \cite{Lange2018Entanglement}. In \EQ{eq:unc}, the first term represents the fluctuations in the particle number difference and the second term represents the fluctuations in the phase difference. In \APP{app:Relation to a number-phase uncertainty}, we show how \EQ{eq:unc} can be connected to the literature on number-phase uncertainties.

It is interesting to note that the term with $(\Delta J_z^s)^2$ in \EQ{eq:unc} has a constant part. The well-known Heisenberg  uncertainty with the canonical $x$ and $p$ operators does not have such a constant. For that uncertainty, when $\va{x}$ is approaching zero, $\va{p}$ is approaching infinity. In our case, the uncertainties of the angular momentum components are bounded from above for a given particle number. Thus, such a constant added to $(\Delta J_z^s)^2$  is needed to obtain a meaningful uncertainty relation with angular momentum components. If $(\Delta J_z^s)^2$ is zero and there were not a constant included in the first term, then an uncertainty could be satisfied only if the right-hand side of the inequality were zero, or the second term on the left-hand side were infinite, which is impossible. Due to the constant term, a state fully polarized in the $z$-direction, having $(\Delta J_z^s)^2=0,$ saturates the relation given in \EQ{eq:unc} with a nonzero value on the right-hand side.

\subsection{Einstein-Podolsky-Rosen criterion}

\label{sec:EPR}

Here, we prove the criterion for steering given in \EQ{eq:stcond}. An important EPR condition with collective spin operators is given as \cite{Cavalcanti2007Uncertainty,Reid2019Quantifying,Dalton2020Tests}
\be
(\Delta_{\inf} J_z\SUPERA)^2 (\Delta_{\inf} J_y\SUPERA)^2\ge \frac 1 4 \ex{J_x\SUPERA}^2 \;, \label{eq:ReidEPR}
\ee
where the right-hand side can be improved considering an inferred value $\ex{J_x\SUPERA}$ based on measurements on party $b$ \cite{Cavalcanti2009Experimental}.
These relations are tailored for states that have a spin almost fully polarized in the $x$-direction. To construct a similar criterion that is also suitable to detect unpolarized states, such as the state given in Eq.~(\ref{eq:Dickestate}), we can use directly Eq.~\eqref{eq:unc}\footnote{Note that there is a relation stronger than \EQ{eq:ReidEPR}, where there is an inferred quantity also on the right-hand side. It can detect also unpolarized states \cite{Cavalcanti2007Uncertainty,Reid2019Quantifying,Dalton2020Tests}.}.

{\it Proof of Observation 2.} We build on the explanation of local hidden state models given in the introduction and the relation given in \EQ{eq:Qa}. Given that one measures an observable $O^b$ on subsystem $b$ and gets outcome $x$, one can associate to subsystem $a$ a positive matrix (i.e., a quantum state that is not normalized) 
\be
\mathcal A^{a}_{x|O^b} = p(x\vert O^b) \sum_\lambda p(\lambda\vert x,O^b) \varrho^a_\lambda.
\label{eq:boundvarinf}
\ee
After all outcomes $x$ are sampled, the whole ensemble of such positive matrices (termed {\it assemblage}) $\{\mathcal A^{a}_{x|O^b}\}$ satisfies $\sum_x \mathcal A^{a}_{x|O^b} = \varrho_{a},$ where $\varrho_{a}=\tr_b(\varrho)$ is the reduced density matrix of $a$. This is consistent with the fact that $a$ and $b$ initially shared a common quantum state $\varrho$.

Now we will derive relations that are valid for states with a local hidden state model. They are uncertainty relations for subsystem $a$ conditioned on measurements on subsystem $b.$ Given the measurement outcomes on $b$, one can try to build estimates  $O^a_{\rm est}$ for all operators $O^a.$  The estimate $O^a_{\rm est}$ is typically a function of the measurement outcomes on subsystem $b.$ 
\be
(\Delta O^a)^2_{\rm inf}=\sum_{x} p(x\vert O^b)   \ex{(O^a-O^a_{{\rm est},x\vert O^b})^2}_{\varrho^a_{x\vert O^b}},
\ee
where we define the normalized density matrix 
\be
\varrho^{a}_{x|O^b} =\frac{\mathcal A^{a}_{x|O^b}}{{\rm Tr}(\mathcal A^{a}_{x|O^b})}= \sum_\lambda p(\lambda\vert x,O^b) \varrho^a_\lambda.\label{eq:rhonorm}
\ee
It is easy to see that the best estimate we obtain for
\be
O^a_{{\rm est},x\vert O^b}=\ex{O^a}_{\varrho^{a}_{x|O^b} }.
\ee
In this case, we obtain 
\be
(\Delta O^a)^2_{\rm inf}=\sum_{x}p(x\vert O^b) \va{O^a}_{\varrho^{a}_{x|O^b}}\label{eq:infvar}
\ee
Due to the concavity of the variance and based on \EQ{eq:rhonorm}, we can write that
\be
\va{O^a}_{\varrho^{a}_{x|O^b}} \ge \sum_\lambda p(\lambda\vert x,O^b) \va{O^a}_{\varrho^a_\lambda}. \label{eq:concavvar}
\ee
Based on \EQ{eq:infvar} and \EQ{eq:concavvar}, for such a local hidden state model the inference variance is lower bounded as \cite{Reid1989Demonstration,Reid2001The,Reid2009Colloquium,Cavalcanti2009Experimental}
\bea
(\Delta O^a)^2_{\rm inf} \ge \sum_{\lambda,x}p(x\vert O^b) p(\lambda\vert x,O^b) \va{O^a}_{\varrho^a_\lambda}  = \sum_\lambda p_\lambda (\Delta O^a)^2_{\varrho^a_\lambda}.\label{eq:DeltaOa}
\eea
On the right-hand side there is the estimation precision reached when we know the local hidden variable.  The inference variance cannot be lower than that.
Note that the above statement is true also for states with a nonzero particle number variance given in \EQ{eq:dmat_varNnonzero_a}.

We will start out from the uncertainty relation given in \EQ{eq:unc}. Let us define the quantities
\be
W:= (\Delta J_x^{a})^2 + (\Delta J_y^{a})^2,\quad\quad L:={\langle (J_x^{a})^2  + (J_y^{a})^2 \rangle}.
\ee
Note that $L\le j_a(j_a+1).$ With this \EQ{eq:unc} can be rewritten for a state $\varrho_a$ in subsystem $a$ as 
\be
\left[ (\Delta J_z^{a})^2_{\varrho_a}+ \frac 1 4 \right] W_{\varrho_{a}}  \ge \frac 1 4 L_{\varrho_a}.\label{eq:uncXY}
\ee
Then, we obtain the following series of inequalities (see Eq.~(34) in \REF{Cavalcanti2009Experimental})
\bea
&&\left[(\Delta_{\rm inf} J_z^{a})^2+\frac{1}{4}\right]\times
[(\Delta_{\rm inf} J_x^{a})^2 + (\Delta_{\rm inf} J_y^{a})^2] \ge \left[\sum_\lambda p_\lambda 
(\Delta J_z^{a})^2_{\varrho^{a}_\lambda}
+\frac{1}{4}\right]\times
\left[\sum_\lambda p_\lambda W_{\varrho^{a}_\lambda} \right]  \nonumber\\
&&\quad\quad\quad\ge \Bigg(\sum_k p_\lambda 
\sqrt{\left[(\Delta J_z^{a})^2_{\varrho^{a}_\lambda}
+\frac{1}{4}\right]
}\times\sqrt{W_{\varrho^{a}_\lambda}}\Bigg)^2\ge \left(\sum_\lambda p_\lambda \frac 1 2 \sqrt{L_{\varrho^{a}_\lambda}}\right)^2\ge\frac{{L_{\varrho^{a}}^2}}{4j_a(j_a+1)}.
 \label{eq:unc_normalized_epr33} 
\eea
In \EQ{eq:unc_normalized_epr33}, for the first inequality we used \EQ{eq:DeltaOa}, for the second inequality we used \EQ{eq:ineq}, and for the third inequality we used \EQ{eq:uncXY}. For the last inequality, we used 
\be
\sum_\lambda p_\lambda \sqrt{x_\lambda} \ge \frac{\sum_\lambda p_\lambda x_\lambda}{\sqrt{x_{\rm max}}},\label{eq:boundsqrt}
\ee
where $0\le x_\lambda\le x_{\rm max}.$ It is based on the fact that $\sqrt{x}\le x/\sqrt{x_{\max}},$ if $0\le x\le x_{\rm max}$ holds.
We say that the state is steerable if \EQ{eq:unc_normalized_epr33} is violated.

In order to understand the bound better in \EQ{eq:boundsqrt}, let us consider the case of two subensembles with probabilities $p_1$ and $p_2=1-p_1,$ and $x_1=0.$ Then, the average of $x_\lambda$
\be
\sum_\lambda p_\lambda x_\lambda=p_2 x_2 \label{eq:aver1}
\ee
is not smaller than 
\be
\left(\sum_\lambda p_\lambda \sqrt{x_\lambda}\right)^2=p_2^2 x_2. \label{eq:aver2}
\ee
If $p_2$ is small then \EQ{eq:aver1} can be much larger than \EQ{eq:aver2}.  Nevertheless, 
\be
\left(\frac{\sum_\lambda p_\lambda x_\lambda}{\sqrt{x_{\max}}}\right)^2=p_2^2 x_2 \times \frac{x_2}{x_{\max}}
\ee
is never larger than the expression given in \EQ{eq:aver2}.
$\qed$

The gain factors $g$ [see \EQ{eq:est}] appear only in the inferred variances, and not in the bound as it happens for entanglement criteria (see e.g., Eq.~(1) of \REF{Fadel2018Spatial}). Therefore, to minimize the criterion it is enough to minimize each inferred variance independently, which can be done by taking the derivative with respect to $g$, and then equating to zero. This straightforward calculation gives
\begin{equation}\label{eq:optg}
g_l = - \dfrac{\text{Cov}(J_l\SUPERA,J_l\SUPERB)}{(\Delta J_l\SUPERB)^2},
\end{equation}
where the covariance is defined as
\be
\text{Cov}(J_l\SUPERA,J_l\SUPERB)=\ex{J_l\SUPERA J_l\SUPERB}-\ex{J_l\SUPERA}\ex{J_l\SUPERB}.\label{eq:cov}
\ee
Based on \EQ{eq:optg}, and calculating the values of the correlations for Dicke states, as we do later in \EQS{eg:Jxn2} and \eqref{eq:Jxzycorr_a_expsplit}, the optimal gain factors for experimentally split Dicke states result to be
\be
g_x=g_y=-N/(N+4),\quad g_z=1\;.\label{eq:optg2}
\ee
For large $N$, these coincide with \EQ{eq:glopt}, which are the values we can consider for simplicity as in typical experiments we have $N\gg 100$.

\subsection{Entanglement criterion}
\label{sec:Entanglement criterion}

Now, let us prove the entanglement criterion given in \EQ{eq:rightJx2Jy2B_withoutsqrtB}.  First, as above we start with a criterion that is suitable for states that are completely polarized in a certain direction. This can be found directly from the argument given by Giovannetti {\it et al.} in \REF{Giovanetti2003Characterizing}. 

{\bf Observation 4.} An entanglement criterion with a bound depending on squared first-moments of the collective spin is
\be
\va{J_z} \left[ \va{ J_x^-}+\va{ J_y^-} \right]\ge \frac 1 4 (\ex{ J_x}^2+\ex{ J_y}^2),
\label{eq:critwithfirstmoments}
\ee
where $ J_l^{-}$ is defined in \EQ{eq:Jlpm}.

{\it Proof.} Let us consider first states with a fixed $j_a$ and $j_b.$ Equation~(16) of \REF{Giovanetti2003Characterizing} shows that for all separable states
\be
\vasq{(U_a \pm U_b)} \vasq{(V_a \pm V_b)} \ge \frac 1 4 (|\ex{C_a}|+|\ex{C_b}|)^2
\label{eq:critwithfirstmoments1}
\ee
holds, where $U_s$ and $V_s$ for $s=a,b$ are operators acting on the two subsystems, and $C_s=i[U_s,V_s].$ Using that $(\vert x\vert+\vert y\vert)^2\ge (x+y)^2$ for any $x,y$ it follows that
\be
\vasq{(U_a + U_b)} \vasq{(V_a - V_b)} \ge \frac 1 4 (\ex{C_a}+\ex{C_b})^2.
\label{eq:critwithfirstmoments11}
\ee
We use \EQ{eq:critwithfirstmoments11} for $U_s=J_z\SUPERN$ and $V_s=J_x\SUPERN$, and for  $U_s=J_z\SUPERN$ and $V_s=J_y\SUPERN$ to obtain two inequalities valid for separable states. Summing these two inequalities, we obtain \EQ{eq:critwithfirstmoments}. 

Let us extend this result to the case of nonzero particle number fluctuations. Using \EQ{eq:ineq} we obtain
\bea
	\sum_{j_a,j_b} p_{j_a,j_b} \va{J_z}_{\varrho_{j_a,j_b}} 
		\sum_{j_a,j_b} p_{j_a,j_b}  \left[ \va{ J_x^-}+\va{ J_y^-} \right]_{\varrho_{j_a,j_b}}	
\ge\frac1 4 \left(\sum_{j_a,j_b} p_{j_a,j_b} \sqrt{\ex{ J_x}^2+\ex{ J_y}^2} \right)^2.\label{eq:uncAA}
\eea
The left-hand side of \EQ{eq:uncAA} is not larger than $\va{J_z}\left[ \va{ J_x^-}+\va{ J_y^-} \right]$ due to the concavity of the variance, while the right-hand side of \EQ{eq:uncAA} is never smaller than the right-hand side of \EQ{eq:critwithfirstmoments} due to the convexity of $f(x,y)=\sqrt{x^2+y^2}.$ $\qed$

The criterion presented in \EQ{eq:critwithfirstmoments} cannot be used for Dicke states, as it has first moments of the angular momentum coordinates on the right-hand side, which are zero for such states. Therefore, we will now derive criteria that work for Dicke states as well.

{\bf Observation 5.} For all separable states 
\be
\left[\va{J_z}+\frac 1 4 \right]\left[ \va{ J_x^-}+\va{ J_y^-} \right] \ge \frac 1 4 
\left\langle \sqrt{ J_x^2+ J_y^2}\right\rangle^2\label{eq:rightJx2Jy2B}
\ee
holds.

{\it Proof.} Let us start from \EQ{eq:critwithfirstmoments}, which is true for all separable states, including states with a fluctuating particle number. We add to both sides the quantity
$\left[\va{ J_x}+\va{ J_y}\right]/4,$ and obtain the inequality 
\bea
\va{J_z} \left[ \va{ J_x^-}+\va{ J_y^-} \right]+\frac 1 4 \left[\va{ J_x}+\va{ J_y}\right]
\ge \frac 1 4 (\ex{ J_x^2}+\ex{ J_y^2}).\label{eq:Jxy2}
\eea
Now, let us consider only product states. For these,
\bea
\va{ J_x}+\va{ J_y}= \va{ J_x^-}+\va{ J_y^-}=\va{J_x^a}+\va{J_x^b}+\va{J_y^a}+\va{J_y^b} \label{eq:Jxyplusminus}
\eea
holds. Based on \EQS{eq:Jxy2} and \eqref{eq:Jxyplusminus}, for product states 
\be
\left[\va{J_z}+\frac 1 4 \right]\left[ \va{ J_x^-}+\va{ J_y^-} \right] \ge \frac 1 4 \ex{ J_x^2+ J_y^2}\label{eq:rightJx2Jy2}
\ee
holds.
\EQL{eq:rightJx2Jy2} is not necessarily true for separable states since the left-hand side is not concave in the state, while the right-hand side is linear.

Let us consider another equation that is true for product states,
namely \EQ{eq:rightJx2Jy2B}, which follows from \EQ{eq:rightJx2Jy2} and from the relation 
\be
\ex{A}\ge\ex{\sqrt{A}}^2 
\ee
that holds for any observable $A$. We claim that \EQ{eq:rightJx2Jy2B} is also true for all separable states. In order to proceed, let us see how to obtain inequalities linear in variances. Let us start from the expression
\bea
x^2 y^2\ge c^2,\label{eq:xyc}
\eea
where $c$ is a constant. From \EQ{eq:xyc} and $\alpha x^2+\beta y^2\ge 2xy\sqrt{\alpha\beta},$ it follows that
\bea
\alpha x^2+\beta y^2& \ge 2c\sqrt{\alpha\beta}, \label{eq:xyc2}
\eea
for any real $\alpha$ and $\beta.$ Thus, the inequality with the product of $x$ and $y$ given in \EQ{eq:xyc} is equivalent to the inequalities with the weighted sum of $x$ and $y$ given in \EQ{eq:xyc2} for all $\alpha$ and $\beta.$
 
Based on these, the inequality in \EQ{eq:rightJx2Jy2B} can be written as
\bea
\alpha \left[\va{J_z}+\frac 1 4 \right] + \beta \left[ \va{ J_x^-}+\va{ J_y^-} \right] \ge\sqrt{\alpha\beta}\left\langle \sqrt{ J_x^2+ J_y^2}\right\rangle.
\label{eq:rightJx2Jy2C}
\eea
So far we know that the inequality in \EQ{eq:rightJx2Jy2C} is true for product states for any choice of the real coefficients $\alpha$'s and $\beta$'s. However, the left-hand side of \EQ{eq:rightJx2Jy2C} is concave in the state, while the right-hand side is linear. Hence, the inequality is valid for all separable states.

One can also see that \EQ{eq:rightJx2Jy2C} corresponds to the tangents of the hyperbola appearing in \EQ{eq:rightJx2Jy2B}. Let us consider all the inequalities that can be obtained from \EQ{eq:rightJx2Jy2C} for all $\alpha$'s and $\beta$'s. They are all fulfilled if and only if  the inequality in \EQ{eq:rightJx2Jy2B} is fulfilled. Hence, \EQ{eq:rightJx2Jy2B} is also valid for all separable states. Note that our proof remains valid for the case of fluctuating number of particles.
$\qed$

\EQL{eq:rightJx2Jy2B} is an entanglement condition that detects the Dicke state given in \EQ{eq:Dickestate} as entangled. However, on the right-hand side we have the expectation value of an operator which is difficult to measure in an experiment. We now show somewhat weaker criteria, in which the operator expectation values on the right-hand side are much easier to measure.

{\it Proof of Observation 3.} Let us consider the case of fixed local particle numbers. We use that for $A\ge 0$ we have 
\be
\ex{\sqrt{A}}\ge \ex{A}/\sqrt{\lambda_{\max}(A)},\label{eq:boundsq}
\ee
where $\lambda_{\max}(A)$ denotes the largest eigenvalue of $A.$ In our case,  the operator $A$ is given by
\be 
A= J_x^2+ J_y^2,
\ee 
and for this operator the maximal eigenvalue is 
\be 
\lambda_{\max}(A)=N(N+2)/4.
\ee  
Hence, the right-hand side of \EQ{eq:rightJx2Jy2B} is never smaller than that of \EQ{eq:rightJx2Jy2B_withoutsqrtB}. 

Let us move towards the case of fluctuating particle numbers. We obtain 
\bea
\ex{\sqrt{A}}=\sum_{j_a,j_b} p_{j_a,j_b} \ex{\sqrt A}_{\varrho_{j_a,j_b}}
\ge \sum_{j_a,j_b} p_{j_a,j_b} \frac{\ex{A}_{\varrho_{j_a,j_b}}}{\sqrt{(j_a+j_b)(j_a+j_b+1)}}
= \left\langle {A}/{\sqrt{N(N+2)}}\right\rangle.
\label{eq:boundsq2}
\eea
For the inequality in \EQ{eq:boundsq2} we used \EQ{eq:boundsq}. For the second equality we just used the definition of an operator acting on a state with varying particle numbers given in \EQ{eq:exvarn}. Hence, the right-hand side of \EQ{eq:rightJx2Jy2B} is never smaller than that of \EQ{eq:rightJx2Jy2B_withoutsqrtB}, even if the variance of the particle number is nonzero. 

Alternatively, Observation 3. can also be proven following the steps similar to that of the proof of Observation 2. The advantage of the proof we presented, that it connects our relation to the criterion of Giovannetti {\it et al.} in \REF{Giovanetti2003Characterizing}. $\qed$

We present another condition, also related to \EQ{eq:rightJx2Jy2B}.

{\bf Observation 6.} The following holds for all separable states
\be
\left[\va{J_z}+\frac 1 4 \right]\left[ \va{ J_x^-}+\va{ J_y^-} \right] \ge 
\frac1 8(\ex{\vert  J_x\vert}+\ex{\vert  J_y\vert})^2.\label{eq:abscrit}
\ee
Here, $\ex{\vert J_l \vert}$ is measured as follows. We measure $J_l$ and compute the absolute value. Then, we measure the average of these absolute values. 

{\it Proof.} The inequality
\be
\sqrt{\frac{x^2+y^2}2}\ge \frac{\sqrt{x^2}+\sqrt{y^2}}2
\ee
holds for any numbers $x$ and $y,$ since the square root is concave.  An analogous relation is  true for matrices, since the square root is matrix concave \cite{Bhatia1997Matrix,Hiai2014Introduction}.  Hence, for operators acting on a state space with given $j_a$ and $j_b$
\be
\sqrt{J_x^2+J_y^2}\ge \frac{\sqrt{J_x^2\vphantom{J_y^2}}+\sqrt{J_y^2}}{\sqrt2}\equiv\frac{|J_x|+|J_y|}{\sqrt2}\label{eq:ineq_abs}
\ee
holds.  Hence, for a state with a fixed subsystem particle numbers 
\be
\left\langle\sqrt{J_x^2+J_y^2}\right\rangle  \ge \frac{\ex{|J_x|+|J_y|} }{\sqrt2}.\label{eq:ineq_abs2}
\ee
Clearly, \EQ{eq:ineq_abs2} is also valid for a system with a fluctuating number of particles. Based on these and on \EQ{eq:rightJx2Jy2B}, the statement of the observation follows. $\qed$

\section{Entanglement detection in the vicinity of the ideal Dicke state}
\label{sec:relevant quantites}

Here, we calculate various relevant operator expectation values for the Dicke state, \eqref{eq:Dickestate}. First, we consider collective quantities, namely, components of the collective angular momentum and their second moments. Then, we divide the particles into two groups, and calculate collective quantities for the two parts. We obtain the correlations of these  quantities, and also the variance of their sum and difference.

These calculations provide an intuition of what kind of quantities we need to use to construct an entanglement condition that detects entanglement close to Dicke states. Clearly, we need quantities for which the Dicke state gives an extremal (i.e., minimal or maximal) or almost extremal value. We will then observe how our criteria are exactly tailored for detecting such states.

\subsection{Collective quantities}

\begin{figure}[t]
\begin{center}
\includegraphics[height=3cm]{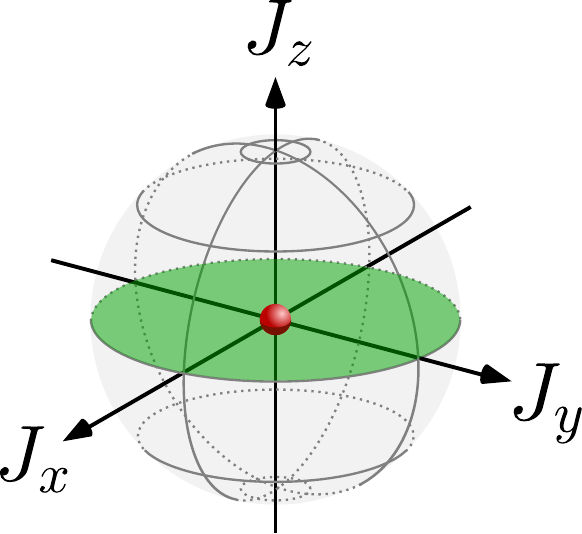}\hskip0.3cm\includegraphics[height=3cm]{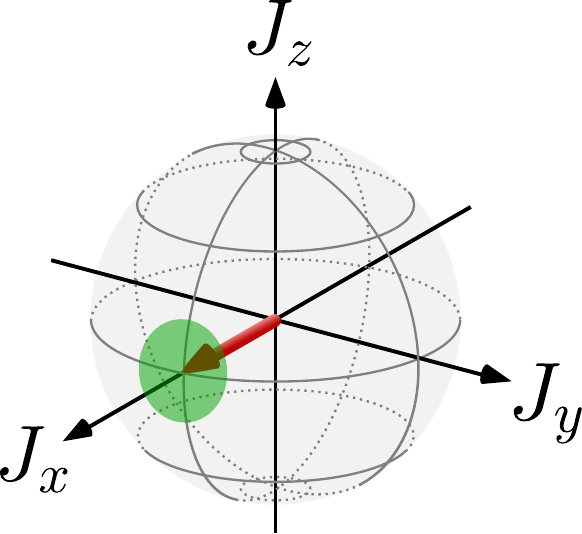}

\hskip0.1cm  (a)\hskip3.5cm (b)

\includegraphics[height=3cm]{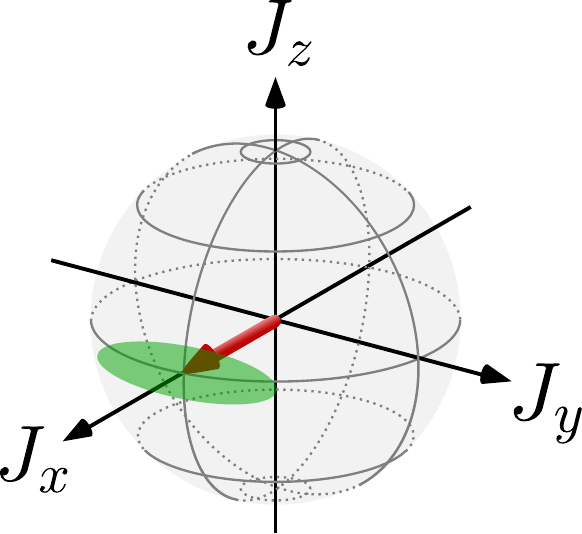}

(c)

\caption{(a) Uncertainty ellipse of a Dicke state given in \EQ{eq:Dickestate}. In the $z$-direction the uncertainty is zero as can be seen from \EQ{eq:secondmomentcollz}, while in the $x$- and $y$-directions it is large, and is given in \EQ{eq:secondmomentcollxy}. The red dot represents the  expectation values of the spin components that are all zero. (b) The same for the state fully polarized in the $x$-direction. The red arrow represents the total spin. (c) The same for the state fully polarized in the $x$-direction and spin-squeezed in the $z$-direction.}
\label{fig:dicke_uncellipse}
\end{center}
\end{figure}

Let us now see the expectation values of the collective observables for the Dicke state of $N$ qubits. First of all, the Dicke state is unpolarized, i.e., $\exs{J_l}=0.$ Moreover, it is symmetric under particle exchange. Hence,
\be
\ex{J_x^2+J_y^2+J_z^2}=\frac N 2 \left (\frac N 2 +1 \right).\label{eq:Jxyz2sym}
\ee 
Since the Dicke state is the eigenstate of $J_z,$ the variance of the $z$-component of the angular momentum is zero
\be
\ex{J_z^2}=0.\label{eq:secondmomentcollz}
\ee
Finally, the Dicke state is symmetric under rotation around the $z$-axis. Hence,
\be
\ex{J_x^m}=\ex{J_y^m} \label{eq:JxJyDicke}
\ee
holds for all powers $m.$ Equations~\eqref{eq:Jxyz2sym}, \eqref{eq:secondmomentcollz} and \eqref{eq:JxJyDicke} lead to
\be
\ex{J_x^2}=\ex{J_y^2}=\frac N 4 \left (\frac N 2 +1 \right).\label{eq:secondmomentcollxy}
\ee
Thus, the uncertainty ellipse of the collective angular momentum is a ``pancake'' which has zero width in the $z$-direction, while in $x$- and $y$-directions its width is large, as it can be seen in \FIG{fig:dicke_uncellipse}(a)~\footnote{Here, the principal axes of the ellipses correspond to the eigenvectors of the covariance matrix $C_{kl}=\tfrac 1 2\ex{J_k J_l + J_l J_k}- \ex{J_k}\ex{J_l}.$ It is also usual to plot the SU(2) Wigner function on the Bloch sphere, which can be displayed as a ring on the equator of the Bloch sphere \cite{Pezze2018Quantum}.}.

\subsection{Bipartite quantities}

In order to detect bipartite entanglement in the Dicke state, we divide the ensemble of $N$ particles into two groups (here without partition noise). The corresponding collective angular momentum components are
\be
J_l\SUPERA=\sum_{n=1}^{N_a} \JLSUPERN, \quad J_l\SUPERB =\sum_{n=N_a+1}^N \JLSUPERN,\label{eq:defJsplit}
\ee
where $l=x,y,z$.
  
Assuming even $N$, we introduce the notation
\be
N_a=\frac N 2+\delta,\quad N_b=\frac N 2-\delta,
\ee
where $-\frac N 2\le \delta \le \frac N 2.$  The value $\delta=0$ corresponds to dividing the ensemble into two equal halves. The corresponding total spins are
\be
j_a=\frac N 4+\frac \delta 2,\quad j_b=\frac N 4-\frac \delta 2.\label{eq:defjlrx}
\ee
Let us calculate now relevant quantities for the Dicke state spilt into two halves. 

For the variance of $J_l\equiv J_l^{a}+J_l^{b},$ we obtain 
\be
\va{J_l} = (\Delta J\SUPERA_l)^2 + (\Delta J\SUPERB_l)^2 + 2\text{Cov}(J_l\SUPERA,J_l\SUPERB),
\ee
while for the variance of $J_l^-$ defined in \EQ{eq:Jlpm}, we obtain 
\be
\va{J_l^-} = (\Delta J\SUPERA_l)^2 + (\Delta J\SUPERB_l)^2 - 2\text{Cov}(J_l\SUPERA,J_l\SUPERB) 
\ee
for $l=x,y,z,$ where the covariance is defined in \EQ{eq:cov}. Clearly, for the Dicke state, for all expectation values in the two parts we have 
\be
\exs{J\SUPERN_l}=0\label{eq:Jln0}
\ee
for $l=x,y,z$ and $\WELL={a,b}.$ 

From \EQ{eq:secondmomentcollxy}, we obtain
\bea
\va{J_x}=\va{J_y}=\frac N 4 \left (\frac N 2 +1 \right) . \label{eq:Jxyplus}
\eea
For large $N$, this is about half of the maximum for any quantum state, which is $N^2/4.$ 

The variances of the difference of the angular momentum components of the two parts are
\bea
\va{J_x^-}=\va{J_y^-}= \frac N 8 \frac{N-2}{N-1} +\frac 1 2 \frac{N}{N-1}\delta^2 \approx \frac N 8+\frac 1 2 \delta^2 , \label{eq:Jxyminus2}
\eea
which is proved in \APP{App:Corr}.
Let us consider the $\delta=0$ case separately, for which
\bea
\va{J_x^-}=\va{J_y^-}&\approx& \frac N 8 . \label{eq:Jxyminus2b}
\eea
As a comparison, for the fully polarized state $\ket{1}^{\otimes N}_z,$ we get 
\be
\va{J_x^-}_{{\rm fp}, z}=\va{J_y^-}_{{\rm fp}, z}= \frac N 4 . \label{eq:Jxyminus2bb}
\ee  
The variance in \EQ{eq:Jxyminus2b}  is smaller than \EQ{eq:Jxyminus2bb}. Note also that  if $\delta$ is nonzero, then  $\va{J_l^-}$ grows rapidly with $\delta,$ as can be seen from \EQ{eq:Jxyminus2}.

Finally, from \EQ{eq:secondmomentcollz}, we obtain for the variance of the sum of the $z$-components 
\bea
\va{J_z}=0.
\eea
That is, the variance of $J_z$ is minimal for Dicke states.

\subsection{Bipartite quantities after splitting in an experiment}

\label{sec:Bipartite quantities splitting in an experiment}

Let us consider now the splitting  that happens in the experiment, which (neglecting the role of particle-particle interactions) mimics a beam-splitter transformation. We will calculate various expectation values for the Dicke state for that case. So far, we assumed that the total particle number $N,$ and also the particle numbers in the two subsystems, $N_a$ and $N_b$ are constants. Consequently, we also assumed that $j_a$ and $j_b$ are constants. In practice, an experiment must be repeated several times in order to obtain sufficient data, and the total particle number is varying from experiment to experiment. Moreover, even if the total particle number $N$ remained constant, when the condensate is split into two parts, we do not obtain two subensembles with exactly $N/2$ particles, but the local particle numbers fluctuate due to partition noise.

If technical noise sources are sufficiently suppressed then the probability of having $N/2+\delta$ particles in subsystem $a$ is given by the binomial formula
\be
p_{\delta}=2^{-N}\binom{N}{N/2+\delta}.\label{eq:px}
\ee
For an ensemble that on average is split equally, the expectation value of $\delta$ is zero. However, the number of particles in the two subsystems fluctuate from experiment to experiment. This is described by the variance of the particle number in the subsystem $a,$ or the variance of $\delta$ which is
\be
{\rm var}(N_a)={\rm var}(\delta)=\exs{\delta^2}=\frac{N}{4} .\label{eq:varx}
\ee
Now we present some quantities after splitting with partition noise, where the expectation values are obtained via an averaging of the expectation values obtained for various $\delta$ values with a weight $p_\delta$ given in \EQ{eq:px}. The derivations are in \APP{app:ExpSplit} (see also Ref.~\cite{Fadel2020Relating}). The local second moments are
\begin{subequations}\label{eg:Jxn2}
\bea
\exs{(J_x\SUPERN)^2}_\delta&=&\exs{(J_y\SUPERN)^2}_\delta=\frac {N(N+4)} {32},\\
\exs{(J_z\SUPERN)^2}_\delta&=& \frac N {16} 
\eea\label{eq:Jxn2Jyn2}
\end{subequations}
for $\WELL={a,b},$ where $\ex{ ... }_\delta$ means averaging over the different $\delta$ values. The correlations are 
\begin{subequations}\label{eq:Jxzycorr_a_expsplit}
\bea
\exs{J_x\SUPERA J_x\SUPERB}_\delta&=&\exs{J_y\SUPERA J_y\SUPERB}_\delta= \frac{N^2}{32},\\ 
\exs{J_z\SUPERA J_z\SUPERB}_\delta&=&-\frac{N}{16}.
\eea\label{eq:Jxzycorr_expsplit}\end{subequations}With these, based on \EQ{eq:optg}, we obtain the $g_l$ coefficients for the Dicke states as in \EQ{eq:optg2}. For simplicity, we will use $\ex{ ... }$ without the $\delta$ subscript in the future. It will be clear from the context, when expectation values and variances are based on an averaging over various $\delta$ values. 

After the correlations, let us write now the collective variances. Based on \EQ{eq:Jxyminus2}, they are obtained as
\bea
\va{J_l^-}= \sum_{\delta=-N/2}^{N/2} p_{\delta} \left(  \frac N 8 \frac{N-2}{N-1} +\frac 1 2 \frac{N}{N-1}\delta^2 \right) 
= \frac N 8 \frac{N-2}{N-1} +\frac 1 2 \frac{N}{N-1}{\rm var}(\delta)=\frac N 4  \label{eq:Jxyminus2c}
\eea
for $l=x,y$
\footnote{$\va{J_l^-}=N/4$ always if a symmetric state is split with partition noise. See \REF{Fadel2020Relating}.}.
Comparing this to \EQ{eq:Jxyminus2b}, we can see that the collective variance became around twice larger due to the partition noise appearing during the splitting of the condensate. This is an unwelcome effect that reduces the quality of the experimental data, and makes the detection of correlations in split states more difficult. 

\subsection{Bipartite quantities with the normalized spin components}
\label{sec:normalization}

Here, we consider a normalization of the collective angular momentum components of the subensembles, since it reduces the unwelcome effects of particle number fluctuations. We calculate some spin expectation values for the Dicke state using a normalization factor, and we show that it cancels the effect of the partition noise. Concretely, we consider the normalized quantities 
\be\label{eq:norm}
\JTILDE_l\SUPERN=J_l\SUPERN/\sqrt{j_s(j_s+1)} 
\ee
for $l=x,y$ and $s=a,b.$ Note that normalized operators appear in the context of fluctuators \cite{Kadar2012SimulatingC,Raggio1989Quantum,Goderis1989Non-commutative,Goderis1989Central}, or in modeling spin systems when normalized angular momentum components perpendicular to the mean spin play the role of the operators $x$ and $p$ of a single mode system \cite{Duan2000Quantum}. However, the angular momentum components are normalized with a different function of the particle number.

Let us also define $\JTILDE_l^{-}=\JTILDE_l\SUPERA - \JTILDE_l\SUPERB$ for $l=x,y$. For their variances on the Dicke state we obtain 
\be
\va{\JTILDE_x^{-}}=\va{\JTILDE_y^{-}}\approx  \frac N {N^2/2+4N-2\delta^2},\label{eq:varDl}
\ee
which is shown in \APP{app:Dj}.

Let us calculate the value for an equal splitting, corresponding to $\delta=0.$ We obtain
\bea
\va{\JTILDE_x^{-}}=\va{\JTILDE_y^{-}}\approx \frac2 N.
\label{eq:Jxyminusnormalized2x0}
\eea
 
We now calculate the value taking into account the variance of $\delta$ while the condensate is split. In this case, the variance of $\delta$ is given in \EQ{eq:varx}, hence $\delta^2\lesssim N/4.$ In \EQ{eq:varDl}, in the denominator we have 
\be
N^2/2\gg 2\delta^2,
\ee
 if $N$ is large. 
Hence, we obtain  \EQ{eq:Jxyminusnormalized2x0}, the same value we would get for an equal splitting with $\delta=0.$ Thus, the fact that the particle number variance after splitting is nonzero did not increase the value of the normalized variance. Remember that the corresponding collective variance without a normalization increased a lot due to the nonzero particle number variance during splitting, as it can be seen in \EQ{eq:Jxyminus2c} when compared to  \EQ{eq:Jxyminus2b}.

Note that we define normalized spin components only for the $x$- and $y$-directions, not in the $z$-direction. The reason is that the variance $\va{J_l^-}$ for $l=x,y$ depends on the splitting ratio $x,$ while $\va{J_z}$ does not.

Finally, let us see now a qualitative argument why the normalization is needed to remove the noise due to splitting with partition noise. For large $N,$ for the Dicke state we have
\be
\va{ J_l^-} = \va{ J_l} - 4 \ex{J_l^a J_l^b}\propto N\label{eq:JlminusMain}
\ee
for $l=x,y.$ In \EQ{eq:JlminusMain}, for the two terms appearing in the difference we have 
\bea
\va{ J_l}&\propto& N^2,\nonumber\\
\ex{J_l^a J_l^b}&\propto& N^2.\label{eq:Dicke_N2}
\eea
Thus, in \EQ{eq:JlminusMain} we have the difference of two quantities that almost cancel each other, and the difference is much smaller than the quantities themselves. This is the reason that \EQ{eq:JlminusMain} is so sensitive to the $\delta,$ as shown in \EQ{eq:Jxyminus2}.

\subsection{Entanglement criteria with normalized quantities}

Next, we present a formulation of our criteria with normalized spin components. The criteria with normalized quantities are more resistant to partition noise, at least in the case of Dicke states (see \SEC{sec:normalization}), and are thus recommended for an experimental use. In short, they can be simply derived repeating the steps described in \SEC{sec:EPR} and \SEC{sec:Entanglement criterion}, after the normalization is inserted in the formulas.

First, we present an EPR steering criterion with normalized quantities.

{\bf Observation 7.} The following EPR steering criterion holds with the normalized quantities 
\bea
\left[(\Delta J_z)^2+\frac{1}{4}\right] \left[  \va{\JTILDE_x^-}+\va{\JTILDE_y^-} \right] \ge \frac 1 4 \ex{(\JTILDE_x\SUPERA)^2+(\JTILDE_y\SUPERA)^2}^{2},
\label{eq:EPR}
\eea
and can be used for Dicke states. A more general form is
\bea
\left([\Delta (J_z\SUPERA-J_{z,{\rm est}}\SUPERA)]^2+\frac{1}{4}\right)\bigg( [\Delta(\JTILDE_{x}\SUPERA-\JTILDE_{x,{\rm est}}\SUPERA)]^2+[\Delta(\JTILDE_{y}\SUPERA-\JTILDE_{y,{\rm est}}\SUPERA)]^2 \bigg) \ge \frac 1 4 \ex{(\JTILDE_x\SUPERA)^2+(\JTILDE_y\SUPERA)^2}^{2}.
\label{eq:EPR2}
\eea
Substituting
\be
\JTILDE_{x,{\rm est}}\SUPERA=\JTILDE_{x}\SUPERB,\quad
\JTILDE_{y,{\rm est}}\SUPERA=\JTILDE_{y}\SUPERB,\quad
J_{z,{\rm est}}\SUPERA=-J_{z}\SUPERB\label{eq:esttilda}
\ee
into \EQ{eq:EPR2}, we obtain \EQ{eq:EPR}.

{\it Proof.} A relation analogous to the one given in \EQ{eq:unc} can be written with normalized operators as 
\be
	\left( (\Delta J_z^s)^2 + \frac 1 4 \right)
		\left[ (\Delta \JTILDE_x^s)^2 + (\Delta \JTILDE_y^s)^2 \right]
	\ge \frac 1 4 (\langle (\JTILDE_x^s)^2 \rangle + \langle (\JTILDE_y^s)^2 \rangle) \label{eq:unc_norm}
\ee
for $s=a,b.$ Following steps similar to those in the proof of Observation 2, we obtain the criteria presented in the Observation. $\qed$

Next, we will present entanglement conditions with normalized quantities.

{\bf Observation 8.} The following inequality with normalized quantities
\be
\left[\va{J_z}+\frac 1 4 \right]\left[ \va{\JTILDE_x^-}+\va{\JTILDE_y^-} \right] \ge \frac 1 {16}
\left\langle\JTILDE_x^2+\JTILDE_y^2\right\rangle^2\label{eq:squarecrit_normalized}
\ee
holds for separable bipartite states. Here, we defined $\JTILDE_l=\JTILDE_l\SUPERA + \JTILDE_l\SUPERB$ for $l=x,y.$

{\it Proof.} A relation analogous to the one given in \EQ{eq:rightJx2Jy2B} can be written with normalized operators as 
\be
\left[\va{J_z}+\frac 1 4 \right]\left[ \va{\JTILDE_x^-}+\va{\JTILDE_y^-} \right] \ge \frac 1 4 
\left\langle \sqrt{\JTILDE_x^2+\JTILDE_y^2}\right\rangle^2.\label{eq:rightJx2Jy2B_norm}
\ee
Then, we have to apply \EQ{eq:boundsq} with $A=\JTILDE_x^2+\JTILDE_y^2$ to obtain an expression that does not have a square root within the expectation value. Here $\lambda_{\max}(A)\le 4,$ where for large $N$ we have $\lambda_{\max}(A)\approx 4.$ $\qed$

{\bf Observation 9.} The following inequality with normalized quantities
\be
\left[\va{J_z}+\frac 1 4 \right]\left[ \va{\JTILDE_x^-}+\va{\JTILDE_y^-} \right] \ge
\frac1 8(\ex{\vert\JTILDE_x\vert}+\ex{\vert\JTILDE_y\vert})^2
\ee
holds for separable bipartite states. 

{\it Proof.} We start from the entanglement criterion \EQ{eq:rightJx2Jy2B_norm}.  Then, we have to apply 
\be
\left\langle\sqrt{\JTILDE_x^2+\JTILDE_y^2}\right\rangle  \ge \frac{\ex{|\JTILDE_x|+|\JTILDE_y|} }{\sqrt2},\label{eq:ineq_abs2b}
\ee
which can be proved similarly to \EQ{eq:ineq_abs}. Then, we obtain an expression that does not have a square root within the expectation value. $\qed$

\begin{table}[t!]
\begin{center}

\begin{tabular}{|l|c|c|c|c|c|c|c|}
\hline
Quantum state & $\vasq{(J_z\SUPERA-J_{z,{\rm est}}\SUPERA)}$ & $V_a$   & $E_a$ & LHS/RHS\\
\hline
$\ket{D_N}$& $0$ & $4/N$ & $1$ &$4/N$ \\
$\ket{+1/2}_z^{\otimes N}$& $0$ & $4/N$  & $4/N$ & $N/4$ \\
$\ket{+1/2}_x^{\otimes N}$& $N/8$ & $2/N$  & $1$ & $1$ \\
\hline
\end{tabular}
\caption{Approximate values for large $N$ of the relevant quantities for the EPR steering criterion given in \EQ{eq:EPR2} for various quantum states. Some of the quantities presented in the table are defined in \EQS{eq:epr1} and \eqref{eq:epr2}, while the EPR condition with those is given in \EQ{eq:epr3}. In the last column, the left-hand side divided by the right-hand side of  \eqref{eq:epr3} is given. A value smaller than one indicates that the state violates the criterion given in \EQ{eq:EPR2}. A value close to one means that the state is close to saturate the relation. 
\label{tab:EPR_entcrit_states}
}
\end{center}
\vskip0.5cm
\begin{center}
\begin{tabular}{|l|c|c|c|c|c|c|c|}
\hline
Quantum state & $\va{J_z}$ & $V$  & $E$ & LHS/RHS\\
\hline
$\ket{D_N}$& $0$ &$4/N$  & $4$ & $1/N$ \\
$\ket{+1/2}_z^{\otimes N}$& $0$ & $8/N$ & $8/N$  & $N/2$\\
$\ket{+1/2}_x^{\otimes N}$& $N/4$ & $4/N$ & $4$ & $1$\\
\hline
\end{tabular}
\caption{Approximate values for large $N$ of the relevant quantities for the entanglement criterion given in \EQ{eq:squarecrit_normalized} for various quantum states. Some of the quantities presented in the table are defined in \EQS{eq:entag1} and \eqref{eq:entag2}, while the entanglement condition with those is given in \eqref{eq:entag3}. In the last column the left-hand side divided by the right-hand side of \eqref{eq:entag3} is given. A value smaller than one indicates that the state violates the criterion given in \EQ{eq:squarecrit_normalized}. A value close to one means that the state is close to saturate the relation.
\label{tab:EPR_entcrit_states2}
}
\end{center}
\end{table}

\section{Application of the entanglement criterion for further quantum states and experimentally prepared Dicke states}

\label{sec:appl}

In this section, we examine how our EPR steering criterion and entanglement criterion work for various ideal quantum states, as well as for a Dicke state prepared experimentally.

\subsection{Testing EPR steering and entanglement for various ideal quantum states}
\label{sec:Examples}

We calculate the relevant normalized quantities for the EPR steering criterion given in \EQ{eq:EPR2} and the entanglement criterion given in \EQ{eq:squarecrit_normalized} for various quantum states.  The results for the EPR condition are shown in \TABLE{tab:EPR_entcrit_states}. The quantities presented in the table are the sum of variances in subsystem $a$
\be
V_a:=[\Delta(\JTILDE_{x}\SUPERA-\JTILDE_{x,{\rm est}}\SUPERA)]^2+[\Delta(\JTILDE_{y}\SUPERA-\JTILDE_{y,{\rm est}}\SUPERA)]^2,\label{eq:epr1}\ee
and the sum of second moments in subsystem $a$
\be E_a:=\ex{(\JTILDE_x^{a})^2}+\ex{(\JTILDE_y^{a})^2}.\label{eq:epr2}\ee
The criterion given in \EQ{eq:EPR2} can be written as 
\be 
\{\vasq{(J_z\SUPERA-J_{z,{\rm est}}\SUPERA)}+1/4\}V_a\ge E_a^{2}/4\label{eq:epr3}
\ee
 with the quantities given in the table.

 The results for the entanglement condition are shown in \TABLE{tab:EPR_entcrit_states2}.
 The quantities presented in the table are the variances of the differences between the two subsystems
\be
V:=\va{\JTILDE_x^-}+\va{\JTILDE_y^-},\label{eq:entag1}
\ee
and the sum of global second moments
\be
E:=\ex{\JTILDE_x^2}+\ex{\JTILDE_y^2},\label{eq:entag2}
\ee
The criterion given in \EQ{eq:squarecrit_normalized} can be written as 
\be
[\va{J_z}+1/4]V\ge E^2/16\label{eq:entag3}
\ee 
with the quantities given in the table.

The Dicke state violates both criteria very strongly. The state with all spins pointing into the $x$-direction, depicted in \FIG{fig:dicke_uncellipse}(b), is close to saturate both relations. All examples are valid when the ensemble is split with partition noise, and also for the case when the ensemble is divided into two equal halves without such noise.  For the Dicke state, based on the arguments presented in this paper, there is not an observable difference  between the two cases for large particle numbers since we used normalized quantities. For the product states, straightforward arguments also show that there is not an observable difference for large particle numbers. Next we discuss, what estimates we used for the various states in the criterion given in \EQS{eq:epr1} and \eqref{eq:epr3}. For the Dicke state, we used the estimates given in \EQ{eq:esttilda}. The other states were product states, hence we used
\be
\JTILDE_{x,{\rm est}}\SUPERA=\JTILDE_{y,{\rm est}}\SUPERA=J_{z,{\rm est}}\SUPERA=0.
\ee

\subsection{Experimental results with Dicke states}
\label{sec:exp}

Next, we test our entanglement criterion with experimental data from \REF{Lange2018Entanglement}. There, a Dicke state of around 8000 atoms was prepared, and bipartite entanglement was detected after splitting. However, in that work a different entanglement criterion was used. Due to the characteristics of the setup, they could measure $J_z\SUPERN$ in the two subsystems. They could also measure 
\be
J_\alpha\SUPERN=J_x\SUPERN \cos(\alpha)+J_y\SUPERN \sin(\alpha),
\ee
where $\WELL=a,b,$ and $\alpha$ is a random angle with a uniform probability distribution between $0$ and $2\pi.$ Note that at a given measurement, the $\alpha$ angle is the same for $J_\alpha\SUPERA$ and $J_\alpha\SUPERB.$

Let us now consider the normalized quantities
\be\label{eq:norm2}
\begin{aligned}
\JTILDE_\alpha\SUPERN &=
J_\alpha\SUPERN/\sqrt{j_s(j_s+1)} ,\\
\JTILDE_\alpha^{\pm}&={\JTILDE}_\alpha\SUPERA\pm{\JTILDE}_\alpha\SUPERB.  
\end{aligned}
\ee
The average $m^{\rm th}$ moment of the angular momentum in the $xy$-plane is defined as
\be
\exs{(\JTILDE_{\bot}^{\pm})^m}=\frac{1}{2\pi}\int_{0}^{2\pi}\exs{(\JTILDE_{\alpha}^{\pm})^m} \ {\rm d}\alpha .
\ee
It is easy to see that the average second moment can be expressed, rather than with an integral, as an average of two operators, namely
\be
\exs{(\JTILDE_{\bot}^{\pm})^2}=\frac{\ex{({\JTILDE_{x}^{\pm})^2+(\JTILDE_{y}^{\pm})^2}}}{2},\label{eq:botxy}
\ee
while the first moment is zero $\exs{\JTILDE_{\bot}^{\pm}}=0.$ \EQL{eq:botxy} can be derived knowing that for the second moments
\be
\frac{1}{2\pi}\int_{0}^{2\pi} (J_\alpha\SUPERN)^2 \ {\rm d}\alpha =\frac{(J_x\SUPERN)^2+(J_y\SUPERN)^2}{2}
\ee
holds, while for the correlations we have
\be
\frac{1}{2\pi}\int_{0}^{2\pi} J_\alpha\SUPERA J_\alpha\SUPERB \ {\rm d}\alpha =\frac{J_x\SUPERA J_x\SUPERB+J_y\SUPERA J_y\SUPERB}{2},
\ee
which can  be proved with straightforward algebra.

With these, the entanglement criterion given in \EQ{eq:squarecrit_normalized} can be rewritten as
\bea
&&\left[ \va{J_z}+\frac{1}{4}\right] \left[2 (\Delta \JTILDE_{\bot}^{-})^2\right]\ge
\frac1 {16} \exs{\JTILDE_{\bot}^2}.
\label{eq:crit_separable2}
\eea

\begin{figure}[t]
\begin{center}
\includegraphics[width=8cm]{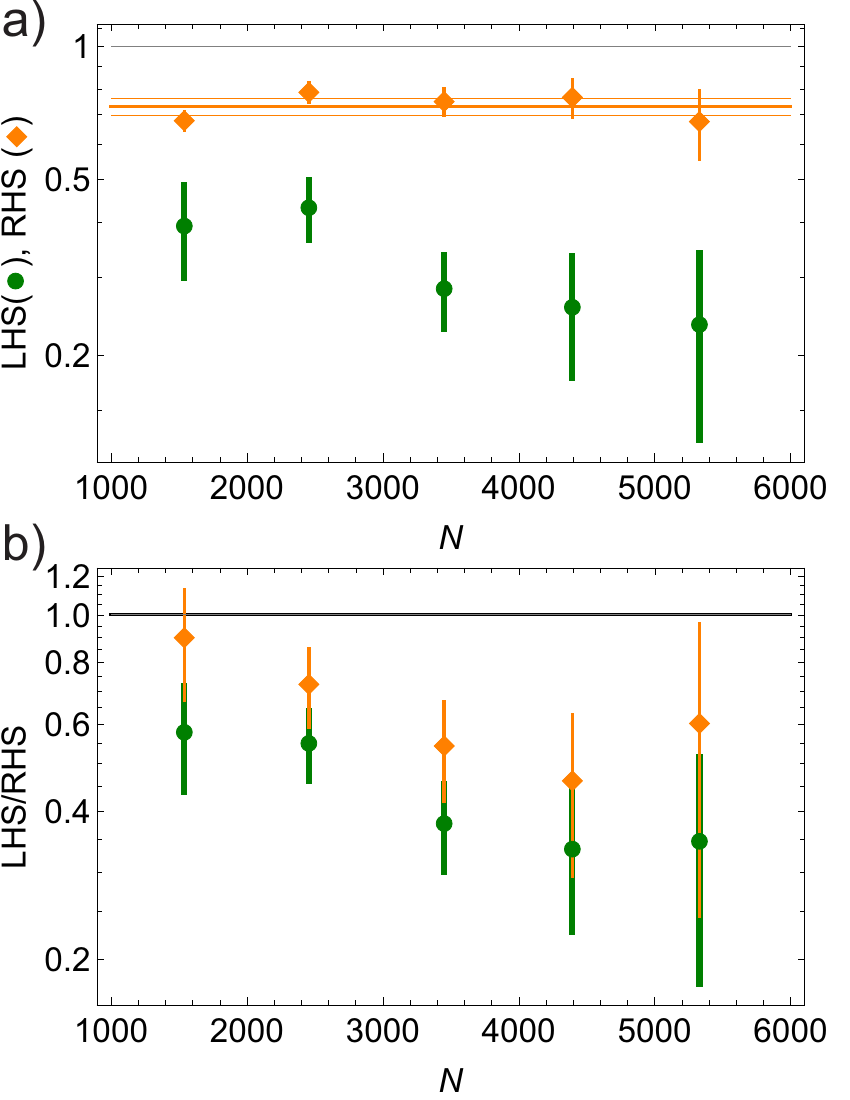}
\caption{Entanglement detection between two spatially separated particle ensembles. (a) Experimental test of the criterion given in \EQ{eq:squarecrit_normalized}. (orange) Right-hand side of the entanglement criterion given in \EQ{eq:squarecrit_normalized}. (green) Left-hand side of \EQ{eq:squarecrit_normalized}. (b) Comparison to the criterion of \REF{Lange2018Entanglement}. (orange) Left-hand side over the right-hand side for the criterion in \REF{Lange2018Entanglement}.  (green) Left-hand side over the right-hand side for the criterion in \EQ{eq:squarecrit_normalized}.}
\label{fig:crit_separable}
\end{center}
\end{figure}

In the experiment described in \REF{Lange2018Entanglement}, they used as point of reference the values obtained for the state fully polarized in the $x$-direction given as
\be
\va{J_z}_{{\rm fp}, x} = \frac N 4,\quad
(\Delta \JTILDE_{\bot}^{-})^2_{{\rm fp},x} = \frac 2 N.
\ee
In \TABLE{tab:EPR_entcrit_states2}, it can be seen that this state is very close to saturate the entanglement criterion \EQ{eq:squarecrit_normalized} for large $N.$ In the experiment described in \REF{Lange2018Entanglement}, $\va{J_z}$  is much smaller than $\va{J_z}_{{\rm fp},x}.$ The reason is that the process that prepares the Dicke state ideally creates  the eigenstate of $J_z$ with eigenvalue $0,$ that is, ideally we have $\va{J_z}=0.$ On the other hand, $\exs{(\JTILDE_{\bot}^{-})^2}$ is larger than $(\Delta \JTILDE_{\bot}^{-})^2_{{\rm fp}, x}$ and the state is close to be symmetric. We plotted the results in \FIG{fig:crit_separable}(a).

We can see that the experimental values violate the condition given in \EQ{eq:crit_separable2}. The inequality presented in this paper is more sensitive than the one presented in  \REF{Lange2018Entanglement}. We can now observe the main characteristics of our method. First, we don't need to assume that the quantum state is fully symmetric. Furthermore, our method can handle particle number variance from experiment to experiment. In practice, the criterion is robust enough to detect entanglement in an actual experiment, where, although the variance of $\va{J_z}$ can be very close to zero, $\va{J_x^-}+\va{J_y^-}$ can be much larger than for the ideal Dicke state. Hence, an entanglement criterion with the product of these two quantities could be more efficient than a criterion with their sum. In \APP{sec:Simple_criteria}, we present some simple entanglement criteria for two spins, which are not yet practical, that help to understand how we developed our criteria.

Let us examine now, how the EPR steering criterion given in \EQ{eq:EPR} could be used in a setup similar to the experiment of \REF{Lange2018Entanglement}. For that, we consider the average $m^{\rm th}$ moment of the angular momentum in the $xy$-plane, defined as
\be
\exs{(\JTILDE_{\bot}\SUPERA)^m}=\frac{1}{2\pi}\int_{0}^{2\pi}\exs{(\JTILDE_{\alpha}\SUPERA)^m} \ {\rm d}\alpha .
\ee
We obtain the relation
\be
\left[ \va{J_z}+\frac{1}{4}\right] \left[2 (\Delta \JTILDE_{\bot}^{-})^2\right]\ge
\frac1 {4} \exs{(\JTILDE_{\bot}\SUPERA)^2}^{2}.
\label{eq:crit_separable2EPR}
\ee

\subsection{Application of the entanglement condition for split spin-squeezed states}
\label{sec:appl2}

\begin{figure}[t]
\begin{center}
\includegraphics{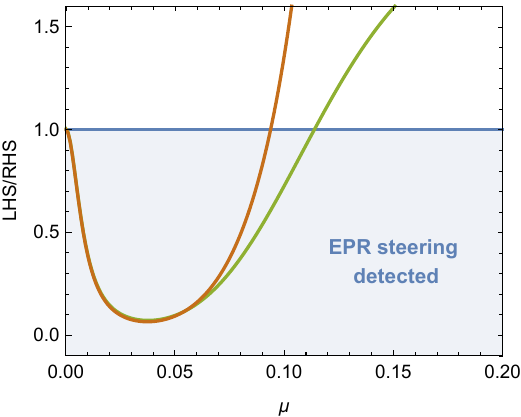}
\caption{Detection of EPR steering in an equally split spin-squeezed state with $N=500$ particles and squeezing strength $\mu,$ for an equally split state (see main text). (green) Left-hand side divided by the right-hand side for the EPR criterion given in \EQ{eq:EPR2}. (brown) Left-hand side divided by the right-hand side for the EPR criterion given in \EQ{eq:ReidEPR}, which is based on the Reid criterion for spins given in \REFS{Cavalcanti2007Uncertainty,Reid2019Quantifying,Dalton2020Tests}. Values below $1$ signal the presence of EPR steering. In both cases the gain factors $g_l$ have been set to their optimal value Eq.~\eqref{eq:optg}.}
\label{fig:oneaxistw}
\end{center}
\end{figure}

\begin{figure}[t]
\begin{center}
\includegraphics{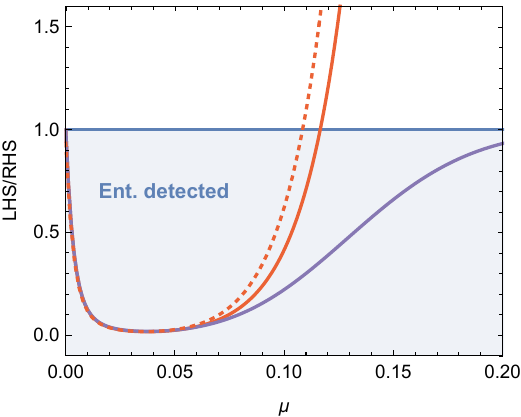} 
\caption{Detection of entanglement in a split spin-squeezed state with $N=500$ particles and squeezing strength $\mu$ (see main text). (purple solid) For an equally split state, left-hand side divided by the right-hand side for the entanglement criterion given in \EQ{eq:squarecrit_normalized}.The curve for splitting with partition noise is not plotted since at this resolution it would overlap completely with the curve corresponding to equal splitting. (red solid)   For an equally split state, left-hand side divided by the right-hand side for the entanglement criterion given in \EQ{eq:critwithfirstmoments11} with  $U_s=J_z\SUPERN,$ $V_s=J_y\SUPERN,$ and $C_s=J_x\SUPERN,$ which is proposed by Giovannetti {\it et al.} in \REF{Giovanetti2003Characterizing}, and analyzed in \REF{Jing2019Split}. Values below $1$ signal the presence of entanglement. (red dashed) The same for splitting with partition noise.}
\label{fig:oneaxistw_ENT}
\end{center}
\end{figure}

From \SEC{sec:Examples} we could see that the state fully polarized in the $x$-direction was close to saturate the entanglement criterion Eq.~\eqref{eq:squarecrit_normalized}. Now, if we start from this state, and apply some suitable interaction between the particles, we can obtain a spin-squeezed state, which has a reduced uncertainty along some direction orthogonal to the mean spin. Such states are entangled, and they are useful, \textit{e.g.}, for quantum metrology, where they outperform separable states \cite{Kitagawa1993Squeezed,Wineland1994Squeezed,Sorensen2001Entanglement}. Spin-squeezed states are now routinely prepared in experiments, and their entanglement has been detected in cold atomic ensembles \cite{Wasilewski2010Quantum,Hald1999Spin,Julsgaard2001Experimental,Hammerer2010Quantum} and in BECs \cite{Gross2012Spin,Esteve2008Squeezing,Riedel2010Atom-chip-based,Ockeloen2013Quantum,Muessel2014Scalable}. In addition, collective measurements allowed to detect and quantify even Bell correlations in such states \cite{Tura2015Nonlocality,Schmied2016Bell,Wagner2017Bell,Baccari2019Bell}. On the other hand, the use of local spin measurements in BECs allowed to detect entanglement and EPR steering between parts of a spin-squeezed state \cite{Fadel2018Spatial,Jing2019Split}.

Here, we show that our entanglement criterion detects bipartite entanglement in split spin-squeezed states.
Such states can be created starting from a spin coherent state polarized along the $x$-direction. Then, we apply a dynamics governed by the one-axis twisting Hamiltonian 
\be
H=\chi J_z^2,
\ee 
where $\chi$ defines the interaction strength. The state can be parametrized during the evolution by the adimensional parameter $\mu=2 \chi t,$ where $t$ is the evolution time. With this parametrization we follow the notation of Ref.~\cite{Kitagawa1993Squeezed}. The state has a reduced spin variance along some direction in the $yz$-plane. By applying a rotation around the $x$-axis, one can orient the state such that the variance is smallest along the $z$-direction. Figures~\ref{fig:dicke_uncellipse}(b) and (c) show the spin and the uncertainty ellipses of the states fully polarized in the $x$-direction and of the state that is also spin-squeezed in the $z$-direction.

Let us now consider a splitting of the state into two halves with equal particle numbers, and investigate correlations between them. The calculations can be simplified since the state is in the symmetric subspace. The resulting bipartite state has been presented in  \REFS{Jing2019Split,Fadel2022Multiparameter}, and it allows us to evaluate analytically \EQ{eq:EPR2}, which is plotted in Fig.~\ref{fig:oneaxistw}. By comparing the violation of the Reid criterion for spins, Eq.~\eqref{eq:ReidEPR}, and of the EPR steering criterion presented in this paper, Eq.~\eqref{eq:EPR2}, we observe that the latter allows us to detect steering for a wider range of $\mu.$ This is the case, because our criterion remains sensitive even when the anti-squeezed component starts to spread around the Bloch sphere.

In \FIG{fig:oneaxistw_ENT}, we show similar curves for the entanglement criterion given in \EQ{eq:squarecrit_normalized}. We compare our approach to the criterion given in \EQ{eq:critwithfirstmoments11} with  $U_s=J_z\SUPERN,$ $V_s=J_y\SUPERN,$ and $C_s=J_x\SUPERN$ which is based on the paper of Giovannetti {\it et al.} given in  \REF{Giovanetti2003Characterizing} (see also \REF{Jing2019Split}). We can see that our criterion allows us to detect spin-squeezed states as entangled even for large $\mu$ values where the state is significantly non-Gaussian.

In Fig.~\ref{fig:oneaxistw_ENT}, we also show the results with the criterion given in \EQ{eq:squarecrit_normalized} for the case of splitting with partition noise. The curve for such case however overlaps with the curve corresponding to equal splitting. Thus, normalization can the help a lot to mitigate the effects of partition noise.

In summary, our criteria for bipartite entanglement developed for the unpolarized Dicke state work also for other classes of states, such as spin-squeezed states. Note that this is  similar to the criteria for multipartite entanglement developed for Dicke states in \REF{Lucke2014Detecting}, that were also applicable for spin-squeezed states \cite{Vitagliano2017Entanglement}.

\section{Discussion}

In this section, we compare our approach to other works about detecting entanglement in bipartite systems. While bipartite entanglement has been detected in many-particle systems in photons \cite{Simon2003Theory,Durkin2002Multiphoton,Eisenberg2004Quantum,Krenn6243Generation} and cold atomic ensembles \cite{Julsgaard2001Experimental}, we consider now in more detail recent works that discuss how to detect bipartite entanglement in  Bose-Einstein condensates of cold atoms.

Some of the proposals are based on detecting bipartite entanglement in split spin-squeezed states, also discussed in \SEC{sec:appl2}. In such systems, the originally fully polarized state of two-state atoms becomes entangled due to an interaction and the variance of a collective spin component, orthogonal to the mean spin, decreases. In \REF{Oudot2017Optimal}, they derive entanglement witnesses using either only first-order or both first- and second-order moments of local collective spin components. In both cases, they derive optimal witnesses for spatially split spin-squeezed states in the presence of local white noise. The correlation-based witness presented is also discussed in \APP{sec:Criterion based on correlations}. In \REF{Oudot2019Bipartite}, it is discussed how to detect bipartite nonlocality in spin systems based on parity measurements, essentially, based on measuring higher-order moments. Experimentally, the observation of EPR correlations between spatially separated modes was achieved in split-squeezed atomic ensembles \cite{Fadel2018Spatial}, using a criterion developed by Giovannetti {\it et al.} in \REF{Giovanetti2003Characterizing}.

Other works consider three-state atoms, originally in the $\ket 0$ state and an interaction in which pairs of atoms from $\ket{0,0}$ tunnel to the $\ket{+1,-1}$ and  $\ket{-1,+1}$ states \cite{Duan2000Squeezing}. In such systems, two operators, different from the angular momentum components, have been presented that have a commutation relation similar to those of the canonical $x$ and $p$ operators, and make it possible to detect spin-nematic squeezing. Based on the ideas above, EPR correlations have been detected experimentally \cite{Kunkel2018Spatially}.

Finally, if after the dynamics mentioned above we project to the subspace with fixed number atoms in states $\ket{+1}$ and $\ket{-1},$ then we obtain a Dicke state given in \EQ{eq:Dickestate}. It is a highly entangled multi-qubit state that has many advantageous properties discussed in \APP{sec:DickeState}. A key fact is that for the Dicke state the expectation value of all the spin components is zero, thus the criteria must contain the second moment of the spin components.  \REF{Lange2018Entanglement} presents an experiment in which entanglement has been detected in such systems with the measurement of collective angular momentum components in the two parties. Our article derives an independent criterion based on general principles, which seems to be stronger when tested on experimental data, as can be seen in \FIG{fig:crit_separable}(b). A key difference is that the present criterion given in \EQ{eq:crit_separable2} includes the term $\exs{(\JTILDE_{\bot}^{+})^2}$ given in \EQ{eq:botxy}, which provides information on how much the bipartite quantum state is in the symmetric subspace. The criterion given in \REF{Lange2018Entanglement} has this type of information only about the subsystems. In the present paper, we also show in detail how the normalization of the spin components cancel some of the partition noise. We also present a criterion for EPR entanglement for Dicke states. We also show that, apart from Dicke states, our criteria work also for split spin-squeezed states.

\section*{Conclusions}
\label{sec:concl}

We presented an uncertainty relation that plays the role of number-phase uncertainty for atomic systems. Based on this uncertainty relation, we described methods to detect Einstein-Podolsky-Rosen steering and bipartite entanglement. Our methods are especially suited for Dicke states, which gave us the motivation for this work, however, they can be also applied to other experimentally relevant states, for instance, split spin-squeezed states. Our methods can handle imperfections, such  a nonzero particle number variance, including the partition noise. They can also handle the problem that the quantum state does not live in a single spatial mode since the populations of the other modes are not fully suppressed. When applied to split nonclassical states, all these conditions need collective spin measurements in the two halves of the system, which are routinely measured in atomic gases and BECs. In the future, it would be interesting to estimate entanglement monotones based on the measured violation of the entanglement criteria ~\cite{FadelVitagliano_2021}.

\acknowledgments
We thank G. Colangelo, O. G\"uhne, P. Hyllus, M. W. Mitchell, and J. Peise for discussions. 
We acknowledge the support of the  EU (COST Action CA15220, QuantERA CEBBEC, QuantERA MENTA, QuantERA QuSiED), the Spanish MCIU (Grant No. PCI2018-092896. No. PCI2022-132947), the Spanish Ministry of Science, Innovation and Universities and the European Regional Development Fund FEDER through Grant No. PGC2018-101355-B-I00 (MCIU/AEI/FEDER, EU) 
and through Grant No. PID2021-126273NB-I00 funded by MCIN/AEI/10.13039/501100011033 and by "ERDF A way of making Europe", the Basque Government (Grant No. IT986-16, No. IT1470-22), and the National Research, Development and Innovation Office NKFIH (Grant No.  K124351, No. KH129601, No. 2019-2.1.7-ERA-NET-2020-00003).  We thank the "Frontline" Research Excellence Programme of the NKFIH (Grant No. KKP133827). We thank Project no. TKP2021-NVA-04, which has been implemented with the support provided by the Ministry of Innovation and Technology of Hungary from the National Research, Development and Innovation Fund, financed under the TKP2021-NVA funding scheme. We thank the Quantum Information National Laboratory of Hungary. G.T. is thankful for a  Bessel Research Award from the Humboldt Foundation.
The work is funded by the Deutsche Forschungsgemeinschaft (DFG, German Research Foundation) under Germany's Excellence Strategy (EXC-2123 QuantumFrontiers 390837967), and through CRC 1227 (DQ-mat), projects A02 and B01.
M. F. was partially supported by the Research Fund of the University of Basel for Excellent Junior Researchers. We also thank the Deutsche Forschungsgemeinschaft (DFG, German Research Foundation, Project No. 447948357), the ERC (Consolidator Grant 683107/TempoQ).
G. V. acknowledges support from the Austrian Science Fund (FWF) through projects ZK 3 (Zukunftskolleg), M 2462-N27 (Lise-Meitner), and P 35810-N (Stand-Alone).

\appendix

\section{The Dicke state as a bipartite state}

\label{sec:DickeState}

In this Appendix, we analyze the entanglement properties of Dicke states given in \EQ{eq:Dickestate}. First of all, the entanglement of Dicke states is quite robust to external noise and imperfections. For instance,  loosing a particle does not make the Dicke state separable. As already mentioned, Dicke states make possible quantum metrology with the maximal, Heisenberg scaling~\cite{Lucke2011Twin}.

The entanglement of formation of $\vert {\rm D}_N \rangle$ is known considering an equal partitioning of the $N$ particles \cite{Stockton2003Characterizing}. For large $N,$ it is given as
\be
E_{\rm F}^{({\rm Dicke})}\approx \log_2(N)/2.\label{eq:Efdicke}
\ee
In comparison, the entanglement of a maximally entangled state of two qudits of dimension $d$ is $\log_2 d.$ Hence, the bipartite entanglement of the Dicke state is close to the entanglement of a maximally entangled state with $d=\sqrt N.$  

Let us observe this in more detail. The Schmidt decomposition of the Dicke state is \cite{Stockton2003Characterizing,Toth2007Detection}
\be
\vert {\rm D}_N \rangle= \sum_{m=0}^N \lambda_m \vert m,N_a \rangle \otimes \vert N/2-m,N_b \rangle ,\label{eq:Dickestate2}
\ee
where the coefficients are
\be
\lambda_m=\binom{N}{N_a}^{-\frac 1 2}\left[\binom{N_a}{m} \binom{N_b}{N/2-m}\right]^{\frac 1 2}
\ee
and the basis states in the two subsystems are
\be
\ket{m,N}=\binom{N}{m}^{-1/2} \sum_k \mathcal{P}_k (\vert 1\rangle^{\otimes m}\vert 0\rangle^{\otimes N-m}).\label{eq:basis}
\ee
The states given in \EQ{eq:basis} are symmetrized superpositions of all product states with $m$ 1's and $N-m$ 0's.

Let us now consider an equal splitting, which means
\be
N_a=N_b=\frac N 2.
\ee
 While the state has $N+1$ nonzero Schmidt coefficients, most of these coefficients are close to zero since for large $N$ the binomial can be approximated with a Gaussian and we obtain
\be
\lambda_m \propto \exp\left[-\frac {(m-N/2)^2} {2(N/4)}\right]. 
\ee
Hence, $\lambda_m$ is  large  if
\be
\vert m-N/2\vert \lesssim \frac {\sqrt N}2.
\ee
This explanation made it easier to understand \EQ{eq:Efdicke} qualitatively.

It is instructive to compare Dicke states to the celebrated Greenberger-Horne-Zeilinger (GHZ) states \cite{Greenberger1990Bells} 
\be
\ket{{\rm GHZ}_N}=\frac 1 {\sqrt 2}\left(\ket{0}^{\otimes N}+\ket{1}^{\otimes N}\right),
\ee
which play a central role in quantum information science. GHZ states have been realized in photonic systems \cite{Bouwmeester1999Observation,Pan2000Experimental,Zhao2003Experimental,Lu2007Experimental,Gao2010Experimental}  and in cold trapped ions \cite{Leibfried2004Toward,Sackett2000Experimental,Monz201114-Qubit}. GHZ states make possible quantum metrology with the maximal, Heisenberg scaling~\cite{Leibfried2004Toward}.

GHZ states have a bipartite entanglement of the two halves of the particles equal to 
\be
E_{\rm F}^{({\rm GHZ})}=\log_2 2=1.
\ee
 Moreover, the entanglement of GHZ states is fragile, and the loss of a single particle can lead to a trivial separable state. 
GHZ states have been created successfully up to 10-20 particles.

\section{Relation to number-phase uncertainty}
\label{app:Relation to a number-phase uncertainty}

In this section, we attempt to connect our derivation to the literature on number-phase uncertainty relations \cite{Carruthers1968Phase,Lynch1995The,LevyLeblond1976Who,Pegg1988Unitary,Barnett1989On,Pegg1989Phase,Vaccaro1990Physical,Luis1993Phase}. If the system is described by a bosonic mode then the number phase relation is written such that the phase operator is related to the annihilation operator $a$ of the mode. Such systems are, for example, light fields, where the superposition of states with different number of photons is detectable. An alternative of the number-phase uncertainty has been used to detect entanglement in two-mode systems \cite{Toth2003Entanglement,Urizar-Lanz2010Number-operator}. An entanglement criterion related to the number-phase uncertainty for two photonic modes, such that one of the modes have few photons, has been considered \cite{Wang2015Strong}. An EPR steering criterion has been presented in such system for two modes \cite{Fadel2020NumberphasePRA}. 

However, in our case we have massive particles. The superposition of quantum states with different particle numbers is not detectable. In other words, the particle number is  preserved by the measurements. In the case of two bosonic modes (e.g., two energy levels), this situation can be handled by mapping the bosonic modes to a spin ensemble using the Schwinger representation \cite{Luis1993Phase,Urizar-Lanz2010Number-operator}
\bea
J_x&=&\frac1 2 (a_1^\dagger a_2+a_2^\dagger a_1),\nonumber\\
J_y&=&\frac i 2 (a_2^\dagger a_1-a_1^\dagger a_2),\nonumber\\
J_z&=&\frac1 2 (a_1^\dagger a_1-a_2^\dagger a_2),
\eea
where $a_k$ are the two modes corresponding to two internal states of the particles. In this case, the role of the annihilation operator $a$ in the number-phase uncertainty relation is played by the operator
\be
J_-=J_x-iJ_y\equiv a_1 a_2^\dagger.
\ee
The operator $J_-$ describes a process in which particles move from one mode to the other,  thus the total particle number does not change.

Based on these, an uncertainty relation resembling a number-phase uncertainty has been obtained between $J_z$ and the operator $E_{12}$ defined via the relation \cite{Luis1993Phase}
\be
J_-=E_{12}\sqrt{J_+J_-},
\ee
where $J_+=J_x+iJ_y,$ which can be further rewritten as
\be
J_-=E_{12}\sqrt{J_x^2+J_y^2+J_z}.
\ee
Since the angular momentum operators are defined with the Schwinger representation for two bosonic modes, the description above is appropriate to describe quantum states of two-state particles with a bosonic symmetry. However, it does not describe general quantum states of $N$ particles that are not symmetric.

We can interpret our approach as one that defines a similar operator
\be
E={J_-}/{\sqrt{\langle J_x^2+J_y^2 \rangle}}=C+iS,
\ee
where the operators $C$ and $S$ are defined as 
\be
C={J_x}/\sqrt{\ex{J_x^2+J_y^2}},\quad
S=-{J_y}/\sqrt{\ex{J_x^2+J_y^2}},
\ee
which are operators roughly corresponding to cosine and sine of the phase. Note that we assume that the quantum state is a multi-qubit state, not necessarily with a bosonic symmetry, and $J_l$ are just the sums of the single-particle spin components as in \EQ{eq:coll}. We can rewrite \EQ{eq:unc} with these operators as
\be
\left[\va{J_z}+\frac 1 4\right]\left[\va{C}+\va{S}\right]\ge \frac 1 4.
\ee

\section{Derivation of \EQ{eq:Jxyminus2}}
\label{App:Corr}

In this Appendix, we calculate various quantities for the Dicke state given in \EQ{eq:Dickestate}. We obtain two-particle and bipartite correlations, and at the end we derive \EQ{eq:Jxyminus2}.

We will now calculate two-particle correlations of the Dicke state. Due to the permutational invariance
\be
\exs{j_l^{(1)} j_l^{(2)}}=\exs{\JLSUPERN j_l^{(m)}}\label{eq:jljl}
\ee
holds for $l=x,y,z$ and for all $n\ne m.$ Hence,
\be
\ex{J_l^2}=\frac{N}{4}+N(N-1)\exs{j_l^{(1)} j_l^{(2)}}.\label{eq:Jl2}
\ee
Then, based on \EQS{eq:secondmomentcollxy} and \eqref{eq:Jl2}
\be
\exs{j_l^{(1)} j_l^{(2)}}=\frac{1}{8}\frac{N}{N-1}\label{eq:corrxxyy}
\ee
for $l=x,y$ while using \EQ{eq:secondmomentcollz} we obtain
\be
\exs{j_z^{(1)} j_z^{(2)}}=-\frac{1}{4(N-1)}.\label{eq:corrzz}
\ee

Using \EQ{eq:defJsplit}, we write the correlations between collective angular momentum component of the two halves with the two-body correlations as
\be
\exs{J_l\SUPERA J_l\SUPERB}=4j_aj_b\exs{j_l^{(1)} j_l^{(2)}}=\left(\frac {N^2}{4}-\delta^2\right)\exs{j_l^{(1)} j_l^{(2)}}\label{eq:JalJbltwoparticle}
\ee
for $l=x,y,z.$ Hence, using \EQS{eq:corrxxyy} and \eqref{eq:corrzz}, we obtain for the correlations between the left and right halves
\begin{subequations}
\bea
\exs{J_x\SUPERA J_x\SUPERB}&=&\exs{J_y\SUPERA J_y\SUPERB}=\left(\frac{N^2}{32}-\frac{\delta^2}{8}\right)\frac{N}{N-1},\label{eq:Jxzycorr_a}\\ 
\exs{J_z\SUPERA J_z\SUPERB}&=&-\left(\frac{N}{16}-\frac{\delta^2}{4N}\right)\frac{N}{N-1}.
\eea\label{eq:Jxzycorr}\end{subequations}
Note that if we measure $J_x$ or $J_y$ on both sides, the measurement results will be {\it correlated} with each other. If we measure $J_z$ on both sides, the measurement results will be {\it anticorrelated} with each other.

It is worth to calculate the maximum of the correlations for any state. We obtain
\be
\vert\exs{J_l\SUPERA J_l\SUPERB}\vert \le j_a  j_b = \frac {N^2} {16} - \frac {\delta^2} 4,\label{eq:maxcorr}
\ee
where the bound is sharp. Note that for large $N$ the correlations given in \EQ{eq:Jxzycorr_a} are only twice less than the absolute maximum given in \EQ{eq:maxcorr} for $\delta=0.$ Thus, the Dicke state has strong correlations both in the $x$- and $y$-directions.

Using \EQS{eq:Jxyplus} and \eqref{eq:Jxzycorr_a}, we obtain
\bea
\va{J_l^-}=\va{J_l}-4\exs{J_l\SUPERA J_l\SUPERB}+ 4\langle J_l\SUPERA\rangle\langle J_l\SUPERB\rangle 
=\frac N 8 \frac{N-2}{N-1} +\frac 1 2 \frac{N}{N-1}\delta^2 \approx \frac N 8+\frac 1 2 \delta^2,\label{eq:Jxyminus2_long}
\eea
which holds for $l=x,y.$ From this, \EQ{eq:Jxyminus2} follows.

\section{Quantities after splitting with partition noise}
\label{app:ExpSplit}

In this Appendix, we calculate various quantities based on the splitting with partition noise characterized by \EQ{eq:varx}.

In order to proceed, we need that
\be
\exs{(J_l\SUPERN)^2}=\left(\sum_{n=1}^{N_a} j_l^{(n)}\right)^2=\frac{j_s}{2}+2j_s(2j_s-1)\exs{j_l^{(1)} j_l^{(2)}},
\label{eq:localsecondmoment}
\ee
for $\WELL=a,b$ and $l=x,y,z,$ where we used that $(\JLSUPERN)^2=\openone/4.$  The two-point correlations are given in \EQS{eq:corrxxyy} and \eqref{eq:corrzz}. Hence, based on \EQ{eq:defjlrx} we obtain
\bea
\exs{(J_l\SUPERA)^2}&=&\frac N 8+\frac \delta 4+\left(\frac N 2+\delta\right)\left(\frac N 2+\delta-1\right)\exs{j_l^{(1)} j_l^{(2)}},\nonumber\\
\exs{(J_l\SUPERB)^2}&=&\frac N 8-\frac \delta 4+\left(\frac N 2-\delta\right)\left(\frac N 2-\delta-1\right)\exs{j_l^{(1)} j_l^{(2)}},\nonumber\\
\eea
for $l=x,y,z.$
Now, considering \EQ{eq:varx} we arrive at \EQ{eq:Jxn2Jyn2}. Based on \EQ{eq:Jxzycorr}, the correlations are obtained as in \EQ{eq:Jxzycorr_expsplit}.

\section{Derivation of the formula for the variance using normalized quantities, given in \EQ{eq:varDl}}
\label{app:Dj}

In this Appendix, we calculate various quantities for the Dicke state given in \EQ{eq:Dickestate}, in order to derive \EQ{eq:varDl}.
Based on \EQ{eq:localsecondmoment}, the normalized second moments are obtained as
\bea
\frac { \exs{(J_l\SUPERN)^2}}{{j_s(j_s+1)}}=\frac1 {2 (j_s+1)}+4\frac{j_s-\frac 1 2}{j_s+1}\exs{j_l^{(1)} j_l^{(2)}}
\approx \frac1 {2 (j_s+1)}+4\exs{j_l^{(1)} j_l^{(2)}}.\label{eq:normmom}
\eea
 for $\WELL=a,b,$ and $l=x,y,$ The two-point correlations are given in \EQ{eq:corrxxyy}.

Let us consider the difference  between the normalized collective angular momentum components of the two subsystems, denoted by $\JTILDE_l^{-}$. For its variance, we obtain
\bea
\va{\JTILDE_l^{-}}
= \frac{\exs{(J_l\SUPERA)^2}}{j_a(j_a+1)}  + \frac{\exs{(J_l\SUPERB)^2}}{j_b(j_b+1)} -8\sqrt{\frac {{j_aj_b}} {{(j_a+1)(j_b+1)} }}\exs{j_l^{(1)} j_l^{(2)}} 
\approx \frac 1 {4(j_a+1)}+\frac 1 {4(j_b+1)},\;\;
\label{eq:Jxyminusnormalized}
\eea
which leads to \EQ{eq:varDl}. Here, in the first equality we used \EQ{eq:Jln0}, which results in $\exs{(\JTILDE_l^{-})}=0$ and we used \EQ{eq:JalJbltwoparticle} giving the correlations. For the second, approximate equality we used the two-body correlations given in \EQ{eq:corrxxyy} and the normalized second moments given in \EQ{eq:normmom}. We also used 
\bea
\sqrt{\frac {j_aj_b} {(j_a+1)(j_b+1)} } =  \sqrt{1-\frac{j_a+j_b+1}{(j_a+1)(j_b+1)}} \approx 1-\frac{j_a+j_b+1}{2(j_a+1)(j_b+1)},
\eea
which is based on that $\sqrt{1-x}\approx 1-x/2$ holds for small $x.$

\section{Basic ideas concerning entanglement detection}
\label{sec:Simple_criteria}

In this section, we present simple criteria that detect the Dicke state given in \EQ{eq:Dickestate} as entangled. These criteria are not yet practical, but will help us to understand the main problems of entanglement detection in such systems.

Next, we assume that the states in the two subsystems are symmetric. Ideally, we can expect this, since the Dicke state is a symmetric multipartite state. If we split it into two subsystems, the quantum state within the subsystems is also symmetric. Hence, the quantum state can be mapped into a quantum state of two large spins.

Based on \EQS{eq:Jxzycorr} and \eqref{eq:varx}, the correlation between the two halves are obtained as in \EQ{eq:Jxzycorr_a_expsplit}. The values given in \EQ{eq:Jxzycorr_a_expsplit} are very close to the values given in \EQ{eq:Jxzycorr}, if we consider large $N$ and substitute $\delta=0.$ Thus, the fact that there is a nonzero particle number variance during the splitting does not change these correlations  a lot compared to the case when the atoms are split into two perfectly equal halves.

\subsection{Criterion based on correlations}
\label{sec:Criterion based on correlations}

Since we have a bipartite system, a first straightforward idea is to detect entanglement based on the correlations between the two parties. For instance, for separable states of two qubits 
\be
\vert \ex{j_x^{(1)}j_x^{(2)}} \vert + \vert \ex{j_y^{(1)}j_y^{(2)}} \vert +  \vert \ex{j_z^{(1)}j_z^{(2)}} \vert  \le 1/4
\ee
holds \cite{Toth2005EntanglementWitnesses,Brukner2004MacroscopicB,Dowling2004Energy}. For two spin-$j$ particles a similar relation can be obtained as 
\be
\vert \ex{j_x^{(1)}j_x^{(2)}} \vert + \vert \ex{j_y^{(1)}j_y^{(2)}} \vert +  \vert \ex{j_z^{(1)}j_z^{(2)}} \vert  \le j^2.
\ee
Let us consider now our Dicke state split into two equal halves. Since the state is in the symmetric subspace, we can map it into a state of two particles of spin $j_a$ and $j_b.$ Then, the entanglement condition reads as
\be
\vert\exs{J_x\SUPERA J_x\SUPERB}\vert+
\vert\exs{J_y\SUPERA J_y\SUPERB}\vert+
\vert\exs{J_z\SUPERA J_z\SUPERB}\vert \le j_a j_b=\frac{N^2}{16}-\frac{\delta^2}{4},\label{eq:corrcrit}
\ee
where we have used \EQ{eq:defjlrx}.
Knowing that in the experiment the variance of $\delta$ is given by \EQ{eq:varx}, we obtain
\be
\vert\exs{J_x\SUPERA J_x\SUPERB}\vert+
\vert\exs{J_y\SUPERA J_y\SUPERB}\vert+
\vert\exs{J_z\SUPERA J_z\SUPERB}\vert \le \frac {N(N-1)} {16},\label{eq:corrcrit2}
\ee
which has been derived in \REF{Oudot2017Optimal}. (Note that \REF{Oudot2017Optimal} presented further interesting results.)

Based on \EQ{eq:Jxzycorr}, for the ideal Dicke state the left-hand side of \EQ{eq:corrcrit}  is 
\be
\frac{N(N+1)}{16}\frac{N}{N-1}-\delta^2\left(\frac 1 4 +\frac 1 {4N}\right)\frac N {N-1}.
\ee
If we consider the splitting for which \EQ{eq:varx} holds, then for  the left-hand side of \EQ{eq:corrcrit}  we obtain
\bea
\frac {N(N+1)} {16}. \label{eq:vD}
\eea
\\\vskip-0.5cm
\noindent The value for the Dicke state given in \EQ{eq:vD} is very close to the bound given on the right-hand side of \EQ{eq:corrcrit2}. Hence, it is difficult to use the criterion for a real experiment. In fact, both the value for Dicke states and the bound is $N^2/16$ in leading order in $N,$ while their difference is proportional to $N$ in leading order in $N.$

\subsection{Criterion based on variances}

\label{sec:Criterion based on variances}

Next, we present simple criteria that use the variances of collective observables. As an introduction, let us consider a well known criterion aimed at detecting singlet states. For separable states of a bipartite system corresponding to the $2j_a: 2j_b$ partition \cite{Toth2004Entanglement},
\be
\va{J_x}+\va{J_y}+\va{J_z}\ge \frac{N}{2}\label{eq:critsinglet}
\ee
holds. For singlet states, the left-hand side of \EQ{eq:critsinglet} is zero, hence they maximally violate the condition.

We would like to apply a similar idea to Dicke states. Hence, we modify the relation
\be
\va{J_x^-}+\va{J_y^-}+\va{J_z} \ge \frac{N}{2}.\label{eq:critdicke}
\ee
The bound for separable states remain the same, however, now the Dicke state violates \EQ{eq:critdicke}. For Dicke states, based on \EQ{eq:Jxyminus2}, the left-hand side is  
\be
2\times\left(\frac{N}{8}\frac{N-2}{N-1} +\frac 1 2 \frac{N}{N-1}\delta^2\right)\approx \frac N 4 + \delta^2.
\ee
For the $\delta=0$ case, representing equal splitting, the Dicke state is detected as entangled. However, the criterion is not very robust. If our Dicke state has ideal correlations in the $x$- and $y$-directions then $\va{J_z}$ can grow up to around $N/4$ such that the Dicke state is still detected. Conversely, if the correlations in the $z$-direction are perfect, then $\va{J_x^-}+\va{J_y^-}$ can grow by around a factor of $2$ to $N/2$ and the state is still detected as entangled. Thus, the criterion tolerates very low noise in $\va{J_x^-}+\va{J_y^-}.$ 

Let us now consider splitting as it is done in experiments, for which \EQ{eq:varx} holds. Based on \EQ{eq:Jxyminus2c}, the left-hand side of \EQ{eq:critdicke} is $N/2$ in this case. That is, the state does not even violate the criterion given in \EQ{eq:critdicke}.

\bibliographystyle{quantum}
\bibliography{twomodedicke_theory80_published}

\begin{thebibliography}{100}

\bibitem{Guhne2009Entanglement}
Otfried G{\"u}hne and G{\'e}za T{\'o}th.
\newblock ``Entanglement detection''.
\newblock
  \href{https://dx.doi.org/https://doi.org/10.1016/j.physrep.2009.02.004}{Phys.
  Rep. {\bf 474}, 1--75}~(2009).

\bibitem{Horodecki2009Quantum}
Ryszard Horodecki, Pawe\l{} Horodecki, Micha\l{} Horodecki, and Karol
  Horodecki.
\newblock ``Quantum entanglement''.
\newblock \href{https://dx.doi.org/10.1103/RevModPhys.81.865}{Rev. Mod. Phys.
  {\bf 81}, 865--942}~(2009).

\bibitem{Friis2019}
Nicolai Friis, Giuseppe Vitagliano, Mehul Malik, and Marcus Huber.
\newblock ``Entanglement certification from theory to experiment''.
\newblock \href{https://dx.doi.org/10.1038/s42254-018-0003-5}{Nat. Rev. Phys.
  {\bf 1}, 72--87}~(2019).

\bibitem{Frerot2023Probing}
Ir\'en\'ee Fr\'erot, Matteo Fadel, and Maciej Lewenstein.
\newblock ``Probing quantum correlations in many-body systems: a review of
  scalable methods''~(2023).
\newblock  \href{http://arxiv.org/abs/2302.00640}{arXiv:2302.00640}.

\bibitem{Pezze2009Entanglement}
Luca Pezz\'e and Augusto Smerzi.
\newblock ``Entanglement, nonlinear dynamics, and the {Heisenberg} limit''.
\newblock \href{https://dx.doi.org/10.1103/PhysRevLett.102.100401}{Phys. Rev.
  Lett. {\bf 102}, 100401}~(2009).

\bibitem{Hyllus2012Fisher}
Philipp Hyllus, Wies\l{}aw Laskowski, Roland Krischek, Christian Schwemmer,
  Witlef Wieczorek, Harald Weinfurter, Luca Pezz\'e, and Augusto Smerzi.
\newblock ``Fisher information and multiparticle entanglement''.
\newblock \href{https://dx.doi.org/10.1103/PhysRevA.85.022321}{Phys. Rev. A
  {\bf 85}, 022321}~(2012).

\bibitem{Toth2012Multipartite}
G\'eza T\'oth.
\newblock ``Multipartite entanglement and high-precision metrology''.
\newblock \href{https://dx.doi.org/10.1103/PhysRevA.85.022322}{Phys. Rev. A
  {\bf 85}, 022322}~(2012).

\bibitem{Raussendorf2001A}
Robert Raussendorf and Hans~J. Briegel.
\newblock ``A one-way quantum computer''.
\newblock \href{https://dx.doi.org/10.1103/PhysRevLett.86.5188}{Phys. Rev.
  Lett. {\bf 86}, 5188--5191}~(2001).

\bibitem{Raussendorf2003Measurement-based}
Robert Raussendorf, Daniel~E. Browne, and Hans~J. Briegel.
\newblock ``Measurement-based quantum computation on cluster states''.
\newblock \href{https://dx.doi.org/10.1103/PhysRevA.68.022312}{Phys. Rev. A
  {\bf 68}, 022312}~(2003).

\bibitem{Gottesman1995Class}
Daniel Gottesman.
\newblock ``Class of quantum error-correcting codes saturating the quantum
  hamming bound''.
\newblock \href{https://dx.doi.org/10.1103/PhysRevA.54.1862}{Phys. Rev. A {\bf
  54}, 1862--1868}~(1996).

\bibitem{Cleve1999How}
Richard Cleve, Daniel Gottesman, and Hoi-Kwong Lo.
\newblock ``How to share a quantum secret''.
\newblock \href{https://dx.doi.org/10.1103/PhysRevLett.83.648}{Phys. Rev. Lett.
  {\bf 83}, 648--651}~(1999).

\bibitem{Curty2004Entanglement}
Marcos Curty, Maciej Lewenstein, and Norbert L\"utkenhaus.
\newblock ``Entanglement as a precondition for secure quantum key
  distribution''.
\newblock \href{https://dx.doi.org/10.1103/PhysRevLett.92.217903}{Phys. Rev.
  Lett. {\bf 92}, 217903}~(2004).

\bibitem{Shor1999Polynomial-Time}
Peter~W. Shor.
\newblock ``Polynomial-time algorithms for prime factorization and discrete
  logarithms on a quantum computer''.
\newblock \href{https://dx.doi.org/10.1137/S0036144598347011}{SIAM Review {\bf
  41}, 303--332}~(1999).

\bibitem{Grover1996A}
L.~K. {Grover}.
\newblock ``{A fast quantum mechanical algorithm for database search}''~(1996).
\newblock
  \href{http://arxiv.org/abs/quant-ph/9605043}{arXiv:quant-ph/9605043}.

\bibitem{Diosi1989Models}
L.~Di\'osi.
\newblock ``Models for universal reduction of macroscopic quantum
  fluctuations''.
\newblock \href{https://dx.doi.org/10.1103/PhysRevA.40.1165}{Phys. Rev. A {\bf
  40}, 1165--1174}~(1989).

\bibitem{Frowis2012Measures}
Florian Fr\"owis and Wolfgang D\"ur.
\newblock ``Measures of macroscopicity for quantum spin systems''.
\newblock \href{https://dx.doi.org/10.1088/1367-2630/14/9/093039}{New J. Phys.
  {\bf 14}, 093039}~(2012).

\bibitem{Sorensen2001Entanglement}
Anders~S. S\o{}rensen and Klaus M\o{}lmer.
\newblock ``Entanglement and extreme spin squeezing''.
\newblock \href{https://dx.doi.org/10.1103/PhysRevLett.86.4431}{Phys. Rev.
  Lett. {\bf 86}, 4431--4434}~(2001).

\bibitem{Toth2007Optimal}
G\'eza T\'oth, Christian Knapp, Otfried G\"uhne, and Hans~J. Briegel.
\newblock ``Optimal spin squeezing inequalities detect bound entanglement in
  spin models''.
\newblock \href{https://dx.doi.org/10.1103/PhysRevLett.99.250405}{Phys. Rev.
  Lett. {\bf 99}, 250405}~(2007).

\bibitem{Toth2009Spin}
G\'eza T\'oth, Christian Knapp, Otfried G\"uhne, and Hans~J. Briegel.
\newblock ``Spin squeezing and entanglement''.
\newblock \href{https://dx.doi.org/10.1103/PhysRevA.79.042334}{Phys. Rev. A
  {\bf 79}, 042334}~(2009).

\bibitem{Lucke2011Twin}
Bernd L{\"u}cke, Manuel Scherer, Jens Kruse, Luca Pezz\'e, Frank Deuretzbacher,
  Phillip Hyllus, Jan Peise, Wolfgang Ertmer, Jan Arlt, Luis Santos, Agosto
  Smerzi, and Carsten Klempt.
\newblock ``Twin matter waves for interferometry beyond the classical limit''.
\newblock \href{https://dx.doi.org/10.1126/science.1208798}{Science {\bf 334},
  773--776}~(2011).

\bibitem{Toth2014Quantum}
G\'eza T\'oth and Iagoba Apellaniz.
\newblock ``Quantum metrology from a quantum information science perspective''.
\newblock \href{https://dx.doi.org/10.1088/1751-8113/47/42/424006}{J. Phys. A:
  Math. Theor. {\bf 47}, 424006}~(2014).

\bibitem{Pezze2018Quantum}
Luca Pezz\`e, Augusto Smerzi, Markus~K. Oberthaler, Roman Schmied, and Philipp
  Treutlein.
\newblock ``Quantum metrology with nonclassical states of atomic ensembles''.
\newblock \href{https://dx.doi.org/10.1103/RevModPhys.90.035005}{Rev. Mod.
  Phys. {\bf 90}, 035005}~(2018).

\bibitem{Duan2011Entanglement}
L.-M. Duan.
\newblock ``Entanglement detection in the vicinity of arbitrary dicke states''.
\newblock \href{https://dx.doi.org/10.1103/PhysRevLett.107.180502}{Phys. Rev.
  Lett. {\bf 107}, 180502}~(2011).

\bibitem{Lucke2014Detecting}
Bernd L\"ucke, Jan Peise, Giuseppe Vitagliano, Jan Arlt, Luis Santos, G\'eza
  T\'oth, and Carsten Klempt.
\newblock ``Detecting multiparticle entanglement of {D}icke states''.
\newblock \href{https://dx.doi.org/10.1103/PhysRevLett.112.155304}{Phys. Rev.
  Lett. {\bf 112}, 155304}~(2014).

\bibitem{Vitagliano2017Entanglement}
Giuseppe Vitagliano, Iagoba Apellaniz, Matthias Kleinmann, Bernd L\"ucke,
  Carsten Klempt, and G\'eza T\'oth.
\newblock ``Entanglement and extreme spin squeezing of unpolarized states''.
\newblock \href{https://dx.doi.org/10.1088/1367-2630/19/1/013027}{New J. Phys.
  {\bf 19}, 013027}~(2017).

\bibitem{Zhang2014Quantum}
Z.~Zhang and L.~M. Duan.
\newblock ``Quantum metrology with dicke squeezed states''.
\newblock \href{https://dx.doi.org/10.1088/1367-2630/16/10/103037}{New J. Phys.
  {\bf 16}, 103037}~(2014).

\bibitem{Apellaniz2015Detecting}
Iagoba Apellaniz, Bernd L\"ucke, Jan Peise, Carsten Klempt, and G\'eza T\'oth.
\newblock ``Detecting metrologically useful entanglement in the vicinity of
  {Dicke} states''.
\newblock
  \href{https://dx.doi.org/http://dx.doi.org/10.1088/1367-2630/17/8/083027}{New
  J. Phys. {\bf 17}, 083027}~(2015).

\bibitem{Apellaniz2017Optimal}
Iagoba Apellaniz, Matthias Kleinmann, Otfried G\"uhne, and G\'eza T\'oth.
\newblock ``Optimal witnessing of the quantum {Fisher} information with few
  measurements''.
\newblock \href{https://dx.doi.org/10.1103/PhysRevA.95.032330}{Phys. Rev. A
  {\bf 95}, 032330}~(2017).

\bibitem{Duan2002Quantum}
L.-M. Duan, J.~I. Cirac, and P.~Zoller.
\newblock ``Quantum entanglement in spinor {Bose-Einstein} condensates''.
\newblock \href{https://dx.doi.org/10.1103/PhysRevA.65.033619}{Phys. Rev. A
  {\bf 65}, 033619}~(2002).

\bibitem{Vitagliano2011Spin}
Giuseppe Vitagliano, Philipp Hyllus, I{\~n}igo~L. Egusquiza, and G\'eza T\'oth.
\newblock ``Spin squeezing inequalities for arbitrary spin''.
\newblock \href{https://dx.doi.org/10.1103/PhysRevLett.107.240502}{Phys. Rev.
  Lett. {\bf 107}, 240502}~(2011).

\bibitem{Vitagliano2014Spin}
Giuseppe Vitagliano, Iagoba Apellaniz, I{\~n}igo~L. Egusquiza, and G\'eza
  T\'oth.
\newblock ``Spin squeezing and entanglement for an arbitrary spin''.
\newblock \href{https://dx.doi.org/10.1103/PhysRevA.89.032307}{Phys. Rev. A
  {\bf 89}, 032307}~(2014).

\bibitem{Toth2010Generation}
G{\'{e}}za T{\'{o}}th and Morgan~W. Mitchell.
\newblock ``Generation of macroscopic singlet states in atomic ensembles''.
\newblock \href{https://dx.doi.org/10.1088/1367-2630/12/5/053007}{New J. Phys.
  {\bf 12}, 053007}~(2010).

\bibitem{Ma2011Quantum}
J.~{Ma}, X.~{Wang}, C.~P. {Sun}, and F.~{Nori}.
\newblock ``{Quantum spin squeezing}''.
\newblock \href{https://dx.doi.org/10.1016/j.physrep.2011.08.003}{Phys. Rep.
  {\bf 509}, 89--165}~(2011).

\bibitem{Kiesel2007Experimental}
N.~Kiesel, C.~Schmid, G.~T\'oth, E.~Solano, and H.~Weinfurter.
\newblock ``Experimental observation of four-photon entangled {D}icke state
  with high fidelity''.
\newblock \href{https://dx.doi.org/10.1103/PhysRevLett.98.063604}{Phys. Rev.
  Lett. {\bf 98}, 063604}~(2007).

\bibitem{Wieczorek2009Experimental}
Witlef Wieczorek, Roland Krischek, Nikolai Kiesel, Patrick Michelberger, G\'eza
  T\'oth, and Harald Weinfurter.
\newblock ``Experimental entanglement of a six-photon symmetric {D}icke
  state''.
\newblock \href{https://dx.doi.org/10.1103/PhysRevLett.103.020504}{Phys. Rev.
  Lett. {\bf 103}, 020504}~(2009).

\bibitem{Prevedel2009Experimental}
R.~Prevedel, G.~Cronenberg, M.~S. Tame, M.~Paternostro, P.~Walther, M.~S. Kim,
  and A.~Zeilinger.
\newblock ``Experimental realization of {Dicke} states of up to six qubits for
  multiparty quantum networking''.
\newblock \href{https://dx.doi.org/10.1103/PhysRevLett.103.020503}{Phys. Rev.
  Lett. {\bf 103}, 020503}~(2009).

\bibitem{Krischek2011Useful}
Roland Krischek, Christian Schwemmer, Witlef Wieczorek, Harald Weinfurter,
  Philipp Hyllus, Luca Pezz\'e, and Augusto Smerzi.
\newblock ``Useful multiparticle entanglement and sub-shot-noise sensitivity in
  experimental phase estimation''.
\newblock \href{https://dx.doi.org/10.1103/PhysRevLett.107.080504}{Phys. Rev.
  Lett. {\bf 107}, 080504}~(2011).

\bibitem{Chiuri2012Experimental}
A.~Chiuri, C.~Greganti, M.~Paternostro, G.~Vallone, and P.~Mataloni.
\newblock ``Experimental quantum networking protocols via four-qubit
  hyperentangled dicke states''.
\newblock \href{https://dx.doi.org/10.1103/PhysRevLett.109.173604}{Phys. Rev.
  Lett. {\bf 109}, 173604}~(2012).

\bibitem{Bohnet2016Quantum}
Justin~G. Bohnet, Brian~C. Sawyer, Joseph~W. Britton, Michael~L. Wall,
  Ana~Maria Rey, Michael Foss-Feig, and John~J. Bollinger.
\newblock ``Quantum spin dynamics and entanglement generation with hundreds of
  trapped ions''.
\newblock \href{https://dx.doi.org/10.1126/science.aad9958}{Science {\bf 352},
  1297--1301}~(2016).

\bibitem{Haffner2005Scalable}
H.~H{\"a}ffner, W.~H{\"a}nsel, C.~F. Roos, J.~Benhelm, M.~Chwalla,
  T.~K{\"o}rber, D.~Rapol, U, M.~Riebe, P.~O. Schmidt, C.~Becher, O.~G\"uhne,
  W.~D\"ur, and R.~Blatt.
\newblock ``Scalable multiparticle entanglement of trapped ions''.
\newblock \href{https://dx.doi.org/10.1038/nature04279}{Nature (London) {\bf
  438}, 643--646}~(2005).

\bibitem{Hume2009Preparation}
D.~B. Hume, C.~W. Chou, T.~Rosenband, and D.~J. Wineland.
\newblock ``Preparation of {Dicke} states in an ion chain''.
\newblock \href{https://dx.doi.org/10.1103/PhysRevA.80.052302}{Phys. Rev. A
  {\bf 80}, 052302}~(2009).

\bibitem{Gross2012Spin}
Christian Gross.
\newblock ``Spin squeezing, entanglement and quantum metrology with
  bose{\textendash}einstein condensates''.
\newblock \href{https://dx.doi.org/10.1088/0953-4075/45/10/103001}{J. Phys. B:
  At. Mol. Opt. Phys. {\bf 45}, 103001}~(2012).

\bibitem{Wasilewski2010Quantum}
W.~Wasilewski, K.~Jensen, H.~Krauter, J.~J. Renema, M.~V. Balabas, and E.~S.
  Polzik.
\newblock ``Quantum noise limited and entanglement-assisted magnetometry''.
\newblock \href{https://dx.doi.org/10.1103/PhysRevLett.104.133601}{Phys. Rev.
  Lett. {\bf 104}, 133601}~(2010).

\bibitem{Fernholz2008Spin}
T.~Fernholz, H.~Krauter, K.~Jensen, J.~F. Sherson, A.~S. S\o{}rensen, and E.~S.
  Polzik.
\newblock ``Spin squeezing of atomic ensembles via nuclear-electronic spin
  entanglement''.
\newblock \href{https://dx.doi.org/10.1103/PhysRevLett.101.073601}{Phys. Rev.
  Lett. {\bf 101}, 073601}~(2008).

\bibitem{Hald1999Spin}
J.~Hald, J.~L. S\o{}rensen, C.~Schori, and E.~S. Polzik.
\newblock ``Spin squeezed atoms: A macroscopic entangled ensemble created by
  light''.
\newblock \href{https://dx.doi.org/10.1103/PhysRevLett.83.1319}{Phys. Rev.
  Lett. {\bf 83}, 1319--1322}~(1999).

\bibitem{Julsgaard2001Experimental}
Brian Julsgaard, Alexander Kozhekin, and Eugene~S Polzik.
\newblock ``Experimental long-lived entanglement of two macroscopic objects''.
\newblock \href{https://dx.doi.org/10.1038/35096524}{Nature (London) {\bf 413},
  400--403}~(2001).

\bibitem{Hammerer2010Quantum}
Klemens Hammerer, Anders~S. S\o{}rensen, and Eugene~S. Polzik.
\newblock ``Quantum interface between light and atomic ensembles''.
\newblock \href{https://dx.doi.org/10.1103/RevModPhys.82.1041}{Rev. Mod. Phys.
  {\bf 82}, 1041--1093}~(2010).

\bibitem{Behbood2014Generation}
N.~Behbood, F.~Martin~Ciurana, G.~Colangelo, M.~Napolitano, G\'eza T\'oth,
  R.~J. Sewell, and M.~W. Mitchell.
\newblock ``Generation of macroscopic singlet states in a cold atomic
  ensemble''.
\newblock \href{https://dx.doi.org/10.1103/PhysRevLett.113.093601}{Phys. Rev.
  Lett. {\bf 113}, 093601}~(2014).

\bibitem{Kong2020Measurement-induced}
Jia Kong, Ricardo Jim{\'e}nez-Mart{\'i}nez, Charikleia Troullinou,
  Vito~Giovanni Lucivero, G{\'e}za T{\'o}th, and Morgan~W. Mitchell.
\newblock ``Measurement-induced, spatially-extended entanglement in a hot,
  strongly-interacting atomic system''.
\newblock \href{https://dx.doi.org/10.1038/s41467-020-15899-1}{Nat. Commun.
  {\bf 11}, 2415}~(2020).

\bibitem{Esteve2008Squeezing}
J.~Esteve, C.~Gross, A.~Weller, S.~Giovanazzi, and M.~K. Oberthaler.
\newblock ``Squeezing and entanglement in a {Bose--Einstein} condensate''.
\newblock
  \href{https://dx.doi.org/http://dx.doi.org/10.1038/nature07332}{Nature
  (London) {\bf 455}, 1216--1219}~(2008).

\bibitem{Gross2010Nonlinear}
Christian Gross, Tilman Zibold, Eike Nicklas, Jerome Esteve, and Markus~K
  Oberthaler.
\newblock ``Nonlinear atom interferometer surpasses classical precision
  limit''.
\newblock
  \href{https://dx.doi.org/http://dx.doi.org/10.1038/nature08919}{Nature
  (London) {\bf 464}, 1165--1169}~(2010).

\bibitem{Riedel2010Atom-chip-based}
Max~F. Riedel, Pascal B{\"o}hi, Yun Li, Theodor~W. H{\"a}nsch, Alice Sinatra,
  and Philipp Treutlein.
\newblock ``Atom-chip-based generation of entanglement for quantum metrology''.
\newblock \href{https://dx.doi.org/10.1038/nature08988}{Nature (London) {\bf
  464}, 1170--1173}~(2010).

\bibitem{Ockeloen2013Quantum}
Caspar~F. Ockeloen, Roman Schmied, Max~F. Riedel, and Philipp Treutlein.
\newblock ``Quantum metrology with a scanning probe atom interferometer''.
\newblock \href{https://dx.doi.org/10.1103/PhysRevLett.111.143001}{Phys. Rev.
  Lett. {\bf 111}, 143001}~(2013).

\bibitem{Muessel2014Scalable}
W.~Muessel, H.~Strobel, D.~Linnemann, D.~B. Hume, and M.~K. Oberthaler.
\newblock ``Scalable spin squeezing for quantum-enhanced magnetometry with
  {Bose-Einstein} condensates''.
\newblock \href{https://dx.doi.org/10.1103/PhysRevLett.113.103004}{Phys. Rev.
  Lett. {\bf 113}, 103004}~(2014).

\bibitem{Hamley2012Spin-nematic}
C.~D. Hamley, C.~S. Gerving, T.~M. Hoang, E.~M. Bookjans, and M.~S. Chapman.
\newblock ``Spin-nematic squeezed vacuum in a quantum gas''.
\newblock \href{https://dx.doi.org/10.1038/nphys2245}{Nat. Phys. {\bf 8},
  305--308}~(2012).

\bibitem{Zou2018Beating}
Yi-Quan Zou, Ling-Na Wu, Qi~Liu, Xin-Yu Luo, Shuai-Feng Guo, Jia-Hao Cao,
  Meng~Khoon Tey, and Li~You.
\newblock ``Beating the classical precision limit with spin-1 {Dicke} states of
  more than 10,000 atoms''.
\newblock \href{https://dx.doi.org/10.1073/pnas.1715105115}{Proc. Natl. Acad.
  Sci. U.S.A. {\bf 115}, 6381--6385}~(2018).

\bibitem{Killoran2014Extracting}
N.~Killoran, M.~Cramer, and M.~B. Plenio.
\newblock ``Extracting entanglement from identical particles''.
\newblock \href{https://dx.doi.org/10.1103/PhysRevLett.112.150501}{Phys. Rev.
  Lett. {\bf 112}, 150501}~(2014).

\bibitem{Krenn6243Generation}
Mario Krenn, Marcus Huber, Robert Fickler, Radek Lapkiewicz, Sven Ramelow, and
  Anton Zeilinger.
\newblock ``Generation and confirmation of a (100 $\times$ 100)-dimensional
  entangled quantum system''.
\newblock \href{https://dx.doi.org/10.1073/pnas.1402365111}{Proc. Natl. Acad.
  Sci. U.S.A. {\bf 111}, 6243--6247}~(2014).

\bibitem{Erker2017Quantifying}
Paul Erker, Mario Krenn, and Marcus Huber.
\newblock ``Quantifying high dimensional entanglement with two mutually
  unbiased bases''.
\newblock \href{https://dx.doi.org/10.22331/q-2017-07-28-22}{{Quantum} {\bf 1},
  22}~(2017).

\bibitem{Lange2018Entanglement}
Karsten Lange, Jan Peise, Bernd L{\"u}cke, Ilka Kruse, Giuseppe Vitagliano,
  Iagoba Apellaniz, Matthias Kleinmann, G{\'e}za T{\'o}th, and Carsten Klempt.
\newblock ``Entanglement between two spatially separated atomic modes''.
\newblock \href{https://dx.doi.org/10.1126/science.aao2035}{Science {\bf 360},
  416--418}~(2018).

\bibitem{Einstein1935Can}
A.~Einstein, B.~Podolsky, and N.~Rosen.
\newblock ``Can quantum-mechanical description of physical reality be
  considered complete?''.
\newblock \href{https://dx.doi.org/10.1103/PhysRev.47.777}{Phys. Rev. {\bf 47},
  777--780}~(1935).

\bibitem{Reid1989Demonstration}
M.~D. Reid.
\newblock ``Demonstration of the {E}instein-{P}odolsky-{R}osen paradox using
  nondegenerate parametric amplification''.
\newblock \href{https://dx.doi.org/10.1103/PhysRevA.40.913}{Phys. Rev. A {\bf
  40}, 913--923}~(1989).

\bibitem{Reid2001The}
M.~D. {Reid}.
\newblock ``{The {E}instein-{P}odolsky-{R}osen Paradox and Entanglement 1:
  Signatures of EPR correlations for continuous variables}''~(2001).
\newblock
  \href{http://arxiv.org/abs/quant-ph/0112038}{arXiv:quant-ph/0112038}.

\bibitem{Reid2009Colloquium}
M.~D. Reid, P.~D. Drummond, W.~P. Bowen, E.~G. Cavalcanti, P.~K. Lam, H.~A.
  Bachor, U.~L. Andersen, and G.~Leuchs.
\newblock ``Colloquium: The {E}instein-{P}odolsky-{R}osen paradox: From
  concepts to applications''.
\newblock \href{https://dx.doi.org/10.1103/RevModPhys.81.1727}{Rev. Mod. Phys.
  {\bf 81}, 1727--1751}~(2009).

\bibitem{Cavalcanti2009Experimental}
E.~G. Cavalcanti, S.~J. Jones, H.~M. Wiseman, and M.~D. Reid.
\newblock ``Experimental criteria for steering and the
  {E}instein-{P}odolsky-{R}osen paradox''.
\newblock \href{https://dx.doi.org/10.1103/PhysRevA.80.032112}{Phys. Rev. A
  {\bf 80}, 032112}~(2009).

\bibitem{Ou1992Realization}
Z.~Y. Ou, S.~F. Pereira, H.~J. Kimble, and K.~C. Peng.
\newblock ``Realization of the {E}instein-{P}odolsky-{R}osen paradox for
  continuous variables''.
\newblock \href{https://dx.doi.org/10.1103/PhysRevLett.68.3663}{Phys. Rev.
  Lett. {\bf 68}, 3663--3666}~(1992).

\bibitem{Peise2015Satisfying}
J.~Peise, I.~Kruse, K.~Lange, B.~L{\"u}cke, L.~Pezz{\`e}, J.~Arlt, W.~Ertmer,
  K.~Hammerer, L.~Santos, A.~Smerzi, and C.~Klempt.
\newblock ``Satisfying the {E}instein-{P}odolsky-{R}osen criterion with massive
  particles''.
\newblock \href{https://dx.doi.org/10.1038/ncomms9984}{Nat. Commun. {\bf 6},
  8984}~(2015).

\bibitem{He2011Einstein}
Q.~Y. He, M.~D. Reid, T.~G. Vaughan, C.~Gross, M.~Oberthaler, and P.~D.
  Drummond.
\newblock ``{E}instein-{P}odolsky-{R}osen entanglement strategies in two-well
  {B}ose-{E}instein condensates''.
\newblock \href{https://dx.doi.org/10.1103/PhysRevLett.106.120405}{Phys. Rev.
  Lett. {\bf 106}, 120405}~(2011).

\bibitem{Jing2019Split}
Yumang Jing, Matteo Fadel, Valentin Ivannikov, and Tim Byrnes.
\newblock ``Split spin-squeezed {B}ose{\textendash}{E}instein condensates''.
\newblock \href{https://dx.doi.org/10.1088/1367-2630/ab3fcf}{New J. Phys. {\bf
  21}, 093038}~(2019).

\bibitem{Guo2021Detecting}
Jiajie Guo, Feng-Xiao Sun, Daoquan Zhu, Manuel Gessner, Qiongyi He, and Matteo
  Fadel.
\newblock ``Detecting {Einstein-Podolsky-Rosen} steering in non-{Gaussian} spin
  states from conditional spin-squeezing parameters''~(2021).
\newblock  \href{http://arxiv.org/abs/2106.13106}{arXiv:2106.13106}.

\bibitem{Fadel2018Spatial}
Matteo Fadel, Tilman Zibold, Boris D{\'e}camps, and Philipp Treutlein.
\newblock ``Spatial entanglement patterns and {E}instein-{P}odolsky-{R}osen
  steering in {B}ose-{E}instein condensates''.
\newblock \href{https://dx.doi.org/10.1126/science.aao1850}{Science {\bf 360},
  409--413}~(2018).

\bibitem{Giovanetti2003Characterizing}
Vittorio Giovannetti, Stefano Mancini, David Vitali, and Paolo Tombesi.
\newblock ``Characterizing the entanglement of bipartite quantum systems''.
\newblock \href{https://dx.doi.org/10.1103/PhysRevA.67.022320}{Phys. Rev. A
  {\bf 67}, 022320}~(2003).

\bibitem{Kunkel2018Spatially}
Philipp Kunkel, Maximilian Pr{\"u}fer, Helmut Strobel, Daniel Linnemann, Anika
  Fr{\"o}lian, Thomas Gasenzer, Martin G{\"a}rttner, and Markus~K Oberthaler.
\newblock ``Spatially distributed multipartite entanglement enables {E}{P}{R}
  steering of atomic clouds''.
\newblock \href{https://dx.doi.org/10.1126/science.aao2254}{Science {\bf 360},
  413--416}~(2018).

\bibitem{FadelVitagliano_2021}
Matteo Fadel, Ayaka Usui, Marcus Huber, Nicolai Friis, and Giuseppe Vitagliano.
\newblock ``Entanglement quantification in atomic ensembles''.
\newblock \href{https://dx.doi.org/10.1103/PhysRevLett.127.010401}{Phys. Rev.
  Lett. {\bf 127}, 010401}~(2021).

\bibitem{Stockton2003Characterizing}
John~K. Stockton, J.~M. Geremia, Andrew~C. Doherty, and Hideo Mabuchi.
\newblock ``Characterizing the entanglement of symmetric many-particle
  spin-$\frac{1}{2}$ systems''.
\newblock \href{https://dx.doi.org/10.1103/PhysRevA.67.022112}{Phys. Rev. A
  {\bf 67}, 022112}~(2003).

\bibitem{Hyllus2010Entanglement}
P.~Hyllus, L.~Pezz\'e, and A.~Smerzi.
\newblock ``Entanglement and sensitivity in precision measurements with states
  of a fluctuating number of particles''.
\newblock \href{https://dx.doi.org/10.1103/PhysRevLett.105.120501}{Phys. Rev.
  Lett. {\bf 105}, 120501}~(2010).

\bibitem{Hyllus2012Entanglement}
Philipp Hyllus, Luca Pezz\'e, Augusto Smerzi, and G\'eza T\'oth.
\newblock ``Entanglement and extreme spin squeezing for a fluctuating number of
  indistinguishable particles''.
\newblock \href{https://dx.doi.org/10.1103/PhysRevA.86.012337}{Phys. Rev. A
  {\bf 86}, 012337}~(2012).

\bibitem{Toth2004Entanglement}
G\'eza T\'oth.
\newblock ``Entanglement detection in optical lattices of bosonic atoms with
  collective measurements''.
\newblock \href{https://dx.doi.org/10.1103/PhysRevA.69.052327}{Phys. Rev. A
  {\bf 69}, 052327}~(2004).

\bibitem{Toth2022Uncertainty}
G\'eza T\'oth and Florian Fr\"owis.
\newblock ``Uncertainty relations with the variance and the quantum fisher
  information based on convex decompositions of density matrices''.
\newblock \href{https://dx.doi.org/10.1103/PhysRevResearch.4.013075}{Phys. Rev.
  Research {\bf 4}, 013075}~(2022).

\bibitem{Cavalcanti2007Uncertainty}
E.~G. Cavalcanti and M.~D. Reid.
\newblock ``Uncertainty relations for the realization of macroscopic quantum
  superpositions and {E}{P}{R} paradoxes''.
\newblock \href{https://dx.doi.org/10.1080/09500340701639623}{J. Mod. Opt. {\bf
  54}, 2373--2380}~(2007).

\bibitem{Reid2019Quantifying}
M.~D. Reid and Q.~Y. He.
\newblock ``Quantifying the mesoscopic nature of {E}instein-{P}odolsky-{R}osen
  nonlocality''.
\newblock \href{https://dx.doi.org/10.1103/PhysRevLett.123.120402}{Phys. Rev.
  Lett. {\bf 123}, 120402}~(2019).

\bibitem{Dalton2020Tests}
B.~J. Dalton, B.~M. Garraway, and M.~D. Reid.
\newblock ``Tests for {E}instein-{P}odolsky-{R}osen steering in two-mode
  systems of identical massive bosons''.
\newblock \href{https://dx.doi.org/10.1103/PhysRevA.101.012117}{Phys. Rev. A
  {\bf 101}, 012117}~(2020).

\bibitem{Bhatia1997Matrix}
R.~Bhatia.
\newblock ``Matrix analysis''.
\newblock Graduate texts in mathematics. Springer, New York. ~(1997).

\bibitem{Hiai2014Introduction}
Fumio Hiai and D{\'e}nes Petz.
\newblock ``Introduction to matrix analysis and applications''.
\newblock Springer Science \& Business Media. ~(2014).

\bibitem{Fadel2020Relating}
Matteo Fadel and Manuel Gessner.
\newblock ``Relating spin squeezing to multipartite entanglement criteria for
  particles and modes''.
\newblock \href{https://dx.doi.org/10.1103/PhysRevA.102.012412}{Phys. Rev. A
  {\bf 102}, 012412}~(2020).

\bibitem{Kadar2012SimulatingC}
Zoltan {Kadar}, Michael {Keyl}, Geza {Toth}, and Zoltan {Zimboras}.
\newblock ``Simulating continuous quantum systems by mean field
  fluctuations''~(2012).
\newblock  \href{http://arxiv.org/abs/1211.2173}{arXiv:1211.2173}.

\bibitem{Raggio1989Quantum}
G.~A. Raggio and R.~F. Werner.
\newblock ``{Quantum statistical mechanics of general mean field systems}''.
\newblock \href{https://dx.doi.org/10.5169/seals-116175}{Helv. Phys. Acta {\bf
  62}, 980--1003}~(1989).

\bibitem{Goderis1989Non-commutative}
D.~Goderis, A.~Verbeure, and P.~Vets.
\newblock ``Non-commutative central limits''.
\newblock \href{https://dx.doi.org/10.1007/BF00341282}{Probab. Theory Relat.
  Fields {\bf 82}, 527--544}~(1989).

\bibitem{Goderis1989Central}
D.~Goderis and P.~Vets.
\newblock ``{Central limit theorem for mixing quantum systems and the
  CCR-algebra of fluctuations}''.
\newblock \href{https://dx.doi.org/cmp/1104178396}{Commun. Math. Phys. {\bf
  122}, 249--265}~(1989).

\bibitem{Duan2000Quantum}
Lu-Ming Duan, J.~I. Cirac, P.~Zoller, and E.~S. Polzik.
\newblock ``Quantum communication between atomic ensembles using coherent
  light''.
\newblock \href{https://dx.doi.org/10.1103/PhysRevLett.85.5643}{Phys. Rev.
  Lett. {\bf 85}, 5643--5646}~(2000).

\bibitem{Kitagawa1993Squeezed}
Masahiro Kitagawa and Masahito Ueda.
\newblock ``Squeezed spin states''.
\newblock \href{https://dx.doi.org/10.1103/PhysRevA.47.5138}{Phys. Rev. A {\bf
  47}, 5138--5143}~(1993).

\bibitem{Wineland1994Squeezed}
D.~J. Wineland, J.~J. Bollinger, W.~M. Itano, and D.~J. Heinzen.
\newblock ``Squeezed atomic states and projection noise in spectroscopy''.
\newblock \href{https://dx.doi.org/10.1103/PhysRevA.50.67}{Phys. Rev. A {\bf
  50}, 67--88}~(1994).

\bibitem{Tura2015Nonlocality}
J.~Tura, R.~Augusiak, A.B. Sainz, B.~L\"ucke, C.~Klempt, M.~Lewenstein, and
  A.~Ac\'{\i}n.
\newblock ``Nonlocality in many-body quantum systems detected with two-body
  correlators''.
\newblock
  \href{https://dx.doi.org/https://doi.org/10.1016/j.aop.2015.07.021}{Ann.
  Phys. {\bf 362}, 370--423}~(2015).

\bibitem{Schmied2016Bell}
Roman Schmied, Jean-Daniel Bancal, Baptiste Allard, Matteo Fadel, Valerio
  Scarani, Philipp Treutlein, and Nicolas Sangouard.
\newblock ``Bell correlations in a {Bose-Einstein} condensate''.
\newblock \href{https://dx.doi.org/10.1126/science.aad8665}{Science {\bf 352},
  441--444}~(2016).

\bibitem{Wagner2017Bell}
Sebastian Wagner, Roman Schmied, Matteo Fadel, Philipp Treutlein, Nicolas
  Sangouard, and Jean-Daniel Bancal.
\newblock ``Bell correlations in a many-body system with finite statistics''.
\newblock \href{https://dx.doi.org/10.1103/PhysRevLett.119.170403}{Phys. Rev.
  Lett. {\bf 119}, 170403}~(2017).

\bibitem{Baccari2019Bell}
F.~Baccari, J.~Tura, M.~Fadel, A.~Aloy, J.-D. Bancal, N.~Sangouard,
  M.~Lewenstein, A.~Ac\'{\i}n, and R.~Augusiak.
\newblock ``Bell correlation depth in many-body systems''.
\newblock \href{https://dx.doi.org/10.1103/PhysRevA.100.022121}{Phys. Rev. A
  {\bf 100}, 022121}~(2019).

\bibitem{Fadel2022Multiparameter}
Matteo Fadel, Benjamin Yadin, Yuping Mao, Tim Byrnes, and Manuel Gessner.
\newblock ``Multiparameter quantum metrology and mode entanglement with
  spatially split nonclassical spin states''~(2022).
\newblock  \href{http://arxiv.org/abs/2201.11081}{arXiv:2201.11081}.

\bibitem{Simon2003Theory}
Christoph Simon and Dik Bouwmeester.
\newblock ``Theory of an entanglement laser''.
\newblock \href{https://dx.doi.org/10.1103/PhysRevLett.91.053601}{Phys. Rev.
  Lett. {\bf 91}, 053601}~(2003).

\bibitem{Durkin2002Multiphoton}
Gabriel~A. Durkin, Christoph Simon, and Dik Bouwmeester.
\newblock ``Multiphoton entanglement concentration and quantum cryptography''.
\newblock \href{https://dx.doi.org/10.1103/PhysRevLett.88.187902}{Phys. Rev.
  Lett. {\bf 88}, 187902}~(2002).

\bibitem{Eisenberg2004Quantum}
H.~S. Eisenberg, G.~Khoury, G.~A. Durkin, C.~Simon, and D.~Bouwmeester.
\newblock ``Quantum entanglement of a large number of photons''.
\newblock \href{https://dx.doi.org/10.1103/PhysRevLett.93.193901}{Phys. Rev.
  Lett. {\bf 93}, 193901}~(2004).

\bibitem{Oudot2017Optimal}
Enky Oudot, Jean-Daniel Bancal, Roman Schmied, Philipp Treutlein, and Nicolas
  Sangouard.
\newblock ``Optimal entanglement witnesses in a split spin-squeezed
  {B}ose-{E}instein condensate''.
\newblock \href{https://dx.doi.org/10.1103/PhysRevA.95.052347}{Phys. Rev. A
  {\bf 95}, 052347}~(2017).

\bibitem{Oudot2019Bipartite}
Enky Oudot, Jean-Daniel Bancal, Pavel Sekatski, and Nicolas Sangouard.
\newblock ``Bipartite nonlocality with a many-body system''.
\newblock \href{https://dx.doi.org/10.1088/1367-2630/ab4c7c}{New J. Phys. {\bf
  21}, 103043}~(2019).

\bibitem{Duan2000Squeezing}
L.-M. Duan, A.~S\o{}rensen, J.~I. Cirac, and P.~Zoller.
\newblock ``Squeezing and entanglement of atomic beams''.
\newblock \href{https://dx.doi.org/10.1103/PhysRevLett.85.3991}{Phys. Rev.
  Lett. {\bf 85}, 3991--3994}~(2000).

\bibitem{Toth2007Detection}
G.~{T{\'o}th}.
\newblock ``{Detection of multipartite entanglement in the vicinity of
  symmetric Dicke states}''.
\newblock \href{https://dx.doi.org/10.1364/JOSAB.24.000275}{J. Opt. Soc. Am. B
  {\bf 24}, 275--282}~(2007).

\bibitem{Greenberger1990Bells}
Daniel~M Greenberger, Michael~A Horne, Abner Shimony, and Anton Zeilinger.
\newblock ``Bell's theorem without inequalities''.
\newblock \href{https://dx.doi.org/10.1119/1.16243}{Am. J. Phys. {\bf 58},
  1131--1143}~(1990).

\bibitem{Bouwmeester1999Observation}
Dik Bouwmeester, Jian-Wei Pan, Matthew Daniell, Harald Weinfurter, and Anton
  Zeilinger.
\newblock ``Observation of three-photon {Greenberger}-{Horne}-{Zeilinger}
  entanglement''.
\newblock \href{https://dx.doi.org/10.1103/PhysRevLett.82.1345}{Phys. Rev.
  Lett. {\bf 82}, 1345}~(1999).

\bibitem{Pan2000Experimental}
Jian-Wei Pan, Dik Bouwmeester, Matthew Daniell, Harald Weinfurter, and Anton
  Zeilinger.
\newblock ``Experimental test of quantum nonlocality in three-photon
  {Greenberger}-{Horne}-{Zeilinger} entanglement''.
\newblock \href{https://dx.doi.org/https://dx.doi.org/10.1038/35000514}{Nature
  (London) {\bf 403}, 515}~(2000).

\bibitem{Zhao2003Experimental}
Zhi Zhao, Tao Yang, Yu-Ao Chen, An-Ning Zhang, Marek \ifmmode~\dot{Z}\else
  \.{Z}\fi{}ukowski, and Jian-Wei Pan.
\newblock ``Experimental violation of local realism by four-photon
  {Greenberger}-{Horne}-{Zeilinger} entanglement''.
\newblock \href{https://dx.doi.org/10.1103/PhysRevLett.91.180401}{Phys. Rev.
  Lett. {\bf 91}, 180401}~(2003).

\bibitem{Lu2007Experimental}
Chao-Yang Lu, Xiao-Qi Zhou, Otfried G{\"u}hne, Wei-Bo Gao, Jin Zhang,
  Zhen-Sheng Yuan, Alexander Goebel, Tao Yang, and Jian-Wei Pan.
\newblock ``Experimental entanglement of six photons in graph states''.
\newblock \href{https://dx.doi.org/10.1038/nphys507}{Nat. Phys. {\bf 3},
  91--95}~(2007).

\bibitem{Gao2010Experimental}
Wei-Bo Gao, Chao-Yang Lu, Xing-Can Yao, Ping Xu, Otfried G{\"u}hne, Alexander
  Goebel, Yu-Ao Chen, Cheng-Zhi Peng, Zeng-Bing Chen, and Jian-Wei Pan.
\newblock ``Experimental demonstration of a hyper-entangled ten-qubit
  {S}chr{\"o}dinger cat state''.
\newblock \href{https://dx.doi.org/10.1038/nphys1603}{Nat. Phys. {\bf 6},
  331--335}~(2010).

\bibitem{Leibfried2004Toward}
D.~Leibfried, M.~D. Barrett, T.~Schaetz, J.~Britton, J.~Chiaverini, W.~M.
  Itano, J.~D. Jost, C.~Langer, and D.~J. Wineland.
\newblock ``Toward {Heisenberg}-limited spectroscopy with multiparticle
  entangled states''.
\newblock
  \href{https://dx.doi.org/http://dx.doi.org/10.1126/science.1097576}{Science
  {\bf 304}, 1476--1478}~(2004).

\bibitem{Sackett2000Experimental}
C.~A. Sackett, D.~Kielpinski, B.~E. King, C.~Langer, V.~Meyer, C.~J. Myatt,
  M.~Rowe, Q.~A. Turchette, W.~M. Itano, D.~J. Wineland, and C.~Monroe.
\newblock ``Experimental entanglement of four particles''.
\newblock \href{https://dx.doi.org/10.1038/35005011}{Nature (London) {\bf 404},
  256--259}~(2000).

\bibitem{Monz201114-Qubit}
Thomas Monz, Philipp Schindler, Julio~T. Barreiro, Michael Chwalla, Daniel
  Nigg, William~A. Coish, Maximilian Harlander, Wolfgang H\"ansel, Markus
  Hennrich, and Rainer Blatt.
\newblock ``14-qubit entanglement: Creation and coherence''.
\newblock \href{https://dx.doi.org/10.1103/PhysRevLett.106.130506}{Phys. Rev.
  Lett. {\bf 106}, 130506}~(2011).

\bibitem{Carruthers1968Phase}
P.~Carruthers and Michael~Martin Nieto.
\newblock ``Phase and angle variables in quantum mechanics''.
\newblock \href{https://dx.doi.org/10.1103/RevModPhys.40.411}{Rev. Mod. Phys.
  {\bf 40}, 411--440}~(1968).

\bibitem{Lynch1995The}
Robert Lynch.
\newblock ``The quantum phase problem: a critical review''.
\newblock
  \href{https://dx.doi.org/https://doi.org/10.1016/0370-1573(94)00095-K}{Phys.
  Rep. {\bf 256}, 367--436}~(1995).

\bibitem{LevyLeblond1976Who}
Jean-Marc L\'evy-Leblond.
\newblock ``Who is afraid of nonhermitian operators? {A} quantum description of
  angle and phase''.
\newblock
  \href{https://dx.doi.org/https://doi.org/10.1016/0003-4916(76)90283-9}{Ann.
  Phys. {\bf 101}, 319--341}~(1976).

\bibitem{Pegg1988Unitary}
D.~T. Pegg and S.~M. Barnett.
\newblock ``Unitary phase operator in quantum mechanics''.
\newblock \href{https://dx.doi.org/10.1209/0295-5075/6/6/002}{Europhys. Lett.
  {\bf 6}, 483--487}~(1988).

\bibitem{Barnett1989On}
S.~M. Barnett and D.~T. Pegg.
\newblock ``On the {H}ermitian optical phase operator''.
\newblock \href{https://dx.doi.org/10.1080/09500348914550021}{J. Mod. Opt. {\bf
  36}, 7--19}~(1989).

\bibitem{Pegg1989Phase}
D.~T. Pegg and S.~M. Barnett.
\newblock ``Phase properties of the quantized single-mode electromagnetic
  field''.
\newblock \href{https://dx.doi.org/10.1103/PhysRevA.39.1665}{Phys. Rev. A {\bf
  39}, 1665--1675}~(1989).

\bibitem{Vaccaro1990Physical}
J.~A. Vaccaro and D.~T. Pegg.
\newblock ``Physical number-phase intelligent and minimum-uncertainty states of
  light''.
\newblock \href{https://dx.doi.org/10.1080/09500349014550041}{J. Mod. Opt. {\bf
  37}, 17--39}~(1990).

\bibitem{Luis1993Phase}
A.~Luis and L.~L. S\'anchez-Soto.
\newblock ``Phase-difference operator''.
\newblock \href{https://dx.doi.org/10.1103/PhysRevA.48.4702}{Phys. Rev. A {\bf
  48}, 4702--4708}~(1993).

\bibitem{Toth2003Entanglement}
G\'eza T\'oth, Christoph Simon, and Juan~Ignacio Cirac.
\newblock ``Entanglement detection based on interference and particle
  counting''.
\newblock \href{https://dx.doi.org/10.1103/PhysRevA.68.062310}{Phys. Rev. A
  {\bf 68}, 062310}~(2003).

\bibitem{Urizar-Lanz2010Number-operator}
I{\~n}igo Urizar-Lanz and G\'eza T\'oth.
\newblock ``Number-operator--annihilation-operator uncertainty as an
  alternative for the number-phase uncertainty relation''.
\newblock \href{https://dx.doi.org/10.1103/PhysRevA.81.052108}{Phys. Rev. A
  {\bf 81}, 052108}~(2010).

\bibitem{Wang2015Strong}
Tian Wang, Hon~Wai Lau, Hamidreza Kaviani, Roohollah Ghobadi, and Christoph
  Simon.
\newblock ``Strong micro-macro entanglement from a weak cross-kerr
  nonlinearity''.
\newblock \href{https://dx.doi.org/10.1103/PhysRevA.92.012316}{Phys. Rev. A
  {\bf 92}, 012316}~(2015).

\bibitem{Fadel2020NumberphasePRA}
Matteo Fadel, Laura Ares, Alfredo Luis, and Qiongyi He.
\newblock ``Number-phase entanglement and {Einstein-Podolsky-Rosen} steering''.
\newblock \href{https://dx.doi.org/10.1103/PhysRevA.101.052117}{Phys. Rev. A
  {\bf 101}, 052117}~(2020).

\bibitem{Toth2005EntanglementWitnesses}
G\'eza T\'oth.
\newblock ``Entanglement witnesses in spin models''.
\newblock \href{https://dx.doi.org/10.1103/PhysRevA.71.010301}{Phys. Rev. A
  {\bf 71}, 010301(R)}~(2005).

\bibitem{Brukner2004MacroscopicB}
C.~{Brukner} and V.~{Vedral}.
\newblock ``{Macroscopic Thermodynamical Witnesses of Quantum
  Entanglement}''~(2004).
\newblock
  \href{http://arxiv.org/abs/quant-ph/0406040}{arXiv:quant-ph/0406040}.

\bibitem{Dowling2004Energy}
Mark~R. Dowling, Andrew~C. Doherty, and Stephen~D. Bartlett.
\newblock ``Energy as an entanglement witness for quantum many-body systems''.
\newblock \href{https://dx.doi.org/10.1103/PhysRevA.70.062113}{Phys. Rev. A
  {\bf 70}, 062113}~(2004).

\end{thebibliography}

\end{document}